\documentclass[10pt]{IEEEtran}

\usepackage[mathscr]{eucal}
\usepackage{eucal}
\usepackage{calrsfs}
\usepackage{graphicx}          % Include this line if your
\usepackage{graphics} % for pdf, bitmapped graphics files
\usepackage{epsfig} % for postscript graphics files
\usepackage{mathptmx} % assumes new font selection scheme installed
\usepackage{times} % assumes new font selection scheme installed
\usepackage{amsmath} % assumes amsmath package installed
\usepackage{amssymb}  % assumes amsmath package installed
\usepackage{ifthen}
\usepackage{booktabs}
\usepackage{algorithm}
\usepackage{algorithmic}
\usepackage{lipsum}
\usepackage{multirow}
\usepackage{setspace}
\usepackage{amsmath}
\usepackage{amscd}
\UseRawInputEncoding
\newtheorem{exmp}{Example}
\newtheorem{theorem}{Theorem}
\newtheorem{lemma}{Lemma}
\newtheorem{remark}{Remark}
\newtheorem{definition}{Definition}
\newtheorem{corollary}{Corollary}
\makeatletter
\newcommand{\thickhline}{%
    \noalign {\ifnum 0=`}\fi \hrule height 1pt
    \futurelet \reserved@a \@xhline
}
\makeatother
\begin{document}
%
% paper title
% can use linebreaks \\ within to get better formatting as desired
\title{State Reconstruction Under Malicious Sensor Attacks }

\author{Wei Liu%<-this % stops a space

\thanks{Wei Liu is with the School of Information and Electronic Engineering, Zhejiang Gongshang University, Hangzhou 310018,
China (e-mail: intervalm@163.com).}% <-this % stops a space
}

% The paper headers
\markboth{}%
{Shell \MakeLowercase{\textit{et al.}}: Bare Demo of IEEEtran.cls for Journals}
% The only time the second header will appear is for the odd numbered pages
% after the title page when using the twoside option.

% make the title area
\maketitle

\vspace{-25pt}
\begin{abstract}
This paper considers the state reconstruction
problem for discrete-time cyber-physical systems when some of the sensors can be arbitrarily corrupted by malicious attacks where the attacked sensors belong to an unknown set.
We first prove that the state is $s$-error correctable if the system under consideration is $s$-sparse
observable where $s$ denotes the maximum number of attacked sensors. Then, two state reconstruction methods are presented where the first method is based on searching elements with
the same value in a set and the second method is developed in terms of searching element
satisfying a given condition. In addition, after establishing and analyzing the conditions that the proposed state reconstruction methods are not effective, we address that it is very hard to prevent
the state reconstruction when either state reconstruction method
proposed in this paper is used. The correctness and effectiveness of the proposed methods are examined via an example of four-dimensional dynamic systems and a real-world example of three-inertia systems.
\end{abstract}
% IEEEtran.cls defaults to using nonbold math in the Abstract.
% This preserves the distinction between vectors and scalars. However,
% if the journal you are submitting to favors bold math in the abstract,
% then you can use LaTeX's standard command \boldmath at the very start
% of the abstract to achieve this. Many IEEE journals frown on math
% in the abstract anyway.

% Note that keywords are not normally used for peerreview papers.
%\vspace{-10pt}
\vspace{-5pt}
\section{Introduction}
\label{}
Cyber-physical systems (CPSs) are dynamical systems in which computational components, physical processes and communication networks are deeply intertwined.
CPSs play a crucial role in improving efficiency, productivity and innovation across a range of sectors, and examples of such systems include autonomous automobile systems,
smart grids, industrial control systems, robotics systems, etc
 \cite{ap1}$-$\cite{ap6}.
However,
CPSs are vulnerable to cyber attacks because they increasingly rely on Internet.
Hence, it is extreme important to enhance the security and resilience of CPSs.
\par
Secure state reconstruction refers to reconstructing the state whose measurement sensors can be corrupted by malicious attacks.
As a fundamental issue in the security and resilience of CPSs, the secure state reconstruction problem has received increasing attention in recent years \cite{nSr1}$-$\cite{Sr9}
where the results presented in \cite{nSr1}$-$\cite{nSr6} did not consider equations of state.
Because the results of secure state reconstruction presented in this paper consider the equation of state, we focus on reviewing the results of secure state reconstruction with equations of state.
\par
By solving a convex optimization problem that minimizes the sum of each row's magnitudes,
a secure state reconstructor was proposed in \cite{Sr1}. Also, the maximum number of
attacked sensors that can be corrected by the state reconstructor was characterized.
To extend the results of \cite{Sr1} in terms of computational efficiency, two algorithms of secure state reconstruct were proposed in \cite{Sr2} using an event-triggered approach
where the number of sensor
measurements is static in the first algorithm and the second algorithm is based on the idea of Luenberger observer. To deal with the combinatorial complexity in a more efficient manner, an algorithm of secure state reconstruct was developed in \cite{Sr3}
using a satisfiability modulo theory approach, which is based on solving an optimization problem with logic, pseudo-Boolean and convex constraints.
As an extension of \cite{Sr3}, bounded noises were added to state equations and measurement sensors in \cite{Sr3a} and \cite{Sr3b}.
In \cite{Sr4}, different from the discrete-time setting in  \cite{Sr1}$-$\cite{Sr3}, the authors considered the secure state reconstruct problem for continuous-time linear systems and an algorithm of secure state reconstruct was proposed using
observability Gramians.
Leveraging the satisfiability modulo theory approach, an observer of secure state
reconstruct is proposed to deal with the exponential complexity  in \cite{Sr5} where the
observer includes two interacting blocks, a theory
solver and a propositional satisfiability solver. In \cite{Sr6}, the authors addressed that
the secure state reconstruction is NP-hard and provided a polynomial-time solution for the problem of
secure state reconstruction. More results about the problem of
secure state reconstruction can be found in \cite{Srm1}$-$\cite{Srm4} and the
references therein. Under standard settings of CPSs without feedback, sensor placement and any priori information of the state,
the previous results of secure state reconstruction suffer from the following two issues:
\begin{itemize}
  \item The state cannot be reconstructed when the number of measurement sensors that are corrupted by malicious attacks is more than half of all measurement sensors. More precisely,
  the state is not reconstructable for\vspace{2pt} systems with $q$ measurement sensors if $s\geq \lceil q/2\rceil$  where $s$ is the  maximum number of measurement sensors that are corrupted by malicious attacks.
  \item The previous results are based on solving optimization problems or designing observers. As a result, the reconstructed values of system state can
  only gradually approach the real state.
\end{itemize}
Hence, to conquer the above mentioned two issues, it is necessary to develop state reconstruction methods so that the state can be accurately reconstructed under proper conditions
even though the number of measurement sensors that are corrupted by malicious attacks is more than half of all measurement sensors, which motivates our research.
\par
This paper studies the secure state reconstruction problem for discrete-time CPSs where some of the sensors can be arbitrarily corrupted by malicious attacks and the attacked sensors belong to an unknown set.
First, using some results presented in this paper, we address that the state is $s$-error correctable after $r$ steps with $r\leq n$ if the system under consideration is $s$-sparse
observable where $n$ denotes the dimension of the state vector. It is worth mentioning that when more than half of all measurement sensors are arbitrarily attacked, the state still can be correctable using the results proposed in this paper as long as
the system under consideration is $s$-sparse observable.
Since the previous results address that the state is not correctable when more than half of all measurement sensors are arbitrarily attacked, the conditions of error correction presented in previous studies
 are more conservative in contrast to that presented in this paper.
 In order to establish the error correction condition, we first prove Theorem \ref{Theorem1} in Appendix \ref{appensT1} where the proof of Theorem \ref{Theorem1} requires using Lemma
\ref{Theorem1}, and then we address the error
correction condition in Remark \ref{Remark3}.
\par
Then, using the proposed result of error correction and some results developed in this paper, two methods of secure state reconstruction are proposed where
the first method uses the strategy of searching elements with the
same value in a set (SESVS) and the second method is based on
searching element satisfying a given condition (SESGC).
 Compared with the previous results, the two proposed secure state reconstruction methods have the following two advantages:
\begin{itemize}
  \item Under proper conditions, the two proposed methods still can reconstruct the state in this case when more than half of all measurement sensors are arbitrarily attacked while the previous results cannot
  reconstruct the state in this case.
  \item The two proposed methods can accurately reconstruct the state, that is, the reconstructed value of system state is equal to the real state.
  However, the previous results reconstruct the state via solving optimization
problems or designing observers, and hence can only obtain approximate solutions.
\end{itemize}
The idea of the first secure state reconstruction method is given in Remark \ref{Remark4} and the main contents of the first method are provided in Theorem \ref{Theorem2}, Corollary \ref{corollary1} and Remark \ref{Remark8}.
The second method is given in Algorithm 1 and a comparison of the two proposed methods is given in Remark \ref{Remark11}.
\par
Moreover, after establishing and analyzing the conditions that the two proposed methods fail to reconstruct the state, we conclude that
 it is very hard to prevent the state reconstruction when either of the two proposed methods
 is used where the corresponding results for the first method are provided in Theorem \ref{Theorem3}, Corollary \ref{corollary2} and Remark \ref{Remark7}, and the corresponding results for the second method are
 given in Theorem \ref{Theorem4} and Remark \ref{Remark10}.
\par
The rest of this paper is organized as follows. In Section \ref{sec2}, the system under consideration is provided, and the corresponding
secure state reconstruction problem is formulated.
In Section \ref{secmr}, a condition of error correction is established and two secure state reconstruction methods are proposed.
In Section \ref{sectionNE}, an example of four-dimensional dynamic systems and a real-world example of three-inertia systems
are presented to test
the correctness and effectiveness
of the two proposed methods. Some concluding remarks are given
in Section \ref{sectionCo}.
\par
    \textit{Notation:}
The $n$-dimensional real Euclidean space is denoted by $\mathbb{R}^n$ and
the cardinality of a set $S$ is denoted by $|S|$. The transpose and inverse of a matrix $A$ are represented by
$A^\textrm{T}$ and $A^{-1}$, respectively.
For a matrix $M$ with $m$ rows and $n$ columns, and a set $S\subseteq \{1,2,\cdots,m\}$, we use $M(S)\in\mathbb{R}^{(m-|S|)\times n}$ to denote the matrix by deleting the rows indexed in $S$.\vspace{2pt}
For example, we take $M=\left[
                          \begin{array}{cc}
                            1 & 0 \\
                            2 & 3 \\
                            4 & 5 \\
                            6 & 6 \\
                          \end{array}
                        \right]
$\vspace{2pt} and $S=\{1,3,4\}$. Then, we have $M(S)=\left[
                          \begin{array}{cc}
                            2 & 3 \\
                          \end{array}
                        \right]$.
For sets $S$ and\vspace{1pt} $Q$ such that $S\subseteq Q$, we use $Q\backslash S$ to stand for the complement\vspace{1pt} of $S$ in $Q$,
namely, $Q\backslash S=\{x\in Q: x\notin S\}$.
For an\vspace{1pt} $n$-dimensional vector $v=(v_1,v_2,\cdots,v_n)^\textrm{T}$, we use $\textsc{Supp}(v)$ to represent the set of nonzero components in $v$, that is,
$ \textsc{Supp}(v)=\big\{i|i\in\{1,2,\cdots,n\},v_i\neq0\big\}$. For a set with $p$ elements, the number of $q$-combinations is denoted by $C_p^q$, that is, $C_p^q=\frac{p!}{q!(p-q)!}$\vspace{2pt} where $!$ denotes the factorial operator.
The ceiling function of a real number $a$ is denoted by $\lceil a\rceil$.
\section{Problem Formulation  \label{sec2}}  Consider the following discrete-time cyber-physical
system
%\newpage
%\noindent
%\vspace{1pt}
\begin{align}x_{k+1}=&Ax_{k}+Bu_k,\label{sxk}\\
y_{k}=&Cx_k+a_{k}, k=0,1,\cdots\label{syk}
\end{align} where \(x_{k}\in\mathbb{R}^n\) is
the system state;
 \(u_k\in\mathbb{R}^p\) is the
exogenous input; \(y_{k}\in\mathbb{R}^{q}\) is the measurement derived from $q$ sensors;
the $i$-th entry of $y_k$, denoted by $y_{k,i}$, is the measurement of sensor $i$; $a_{k}\in\mathbb{R}^{q}$ is the attacks injected by
an adversary in the sensors; the $i$-th entry of $a_{k}$, denoted by $a_{k,i}$, is the attack on the $i$-th sensor; $A$, $B$ and $C$ are matrices of appropriate dimensions.
\par
If sensor $i$ is attacked by an adversary, $a_{k,i}$ can take arbitrary real number. Otherwise, that is, sensor $i$ is not attacked, $a_{k,i}$ is equal to zero.
It is assumed that
the set of sensors the adversary can attack is unknown and does not change over time.
Also, we assume that the maximum number of sensors that are attacked is
$s$, namely, $\|a_{k}\|_{0}\leq s$. Noting that the maximum number of attacked sensors is
$s$ and the set of attacked sensors
does not change over time, there exists a set
$\Gamma$ with $|\Gamma|= s$ such that \textsc{Supp}$(a_k)\subseteq\Gamma$ where $\Gamma$ is referred to as the index
set of attacked sensors in the remaining of this paper.
\par
The main aim of this paper is to study the secure state reconstruction problem, that is, to
reconstruct the state for the system under consideration when some of the
sensors can be arbitrarily corrupted by malicious attacks.
\section{Secure State Reconstruction \label{secmr}}
In this section, we study the secure state reconstruction problem for the system under consideration. In section
\ref{secmra}, the error correction for the
system under consideration is investigated. In section
\ref{secmrb}, we study the secure state reconstruction problem based on SESVS. In section
\ref{secmrc},  the secure state reconstruction problem is investigated based on SESGC.
\subsection{Error Correction \label{secmra}}
\begin{definition}\cite{Sr2}
The system under consideration is said to be $s$-sparse
observable if $(A,C(\Gamma))$ is observable for any $\Gamma$ satisfying $\Gamma\subseteq\{1,2,\cdots,q\}$ and $|\Gamma|= s$.
\label{definition1}
\end{definition}
\par
The following definition comes from \cite{Sr1} and we use $|\Gamma|= s$ instead of $|\Gamma|\leq s$ by noting that the index set $\Gamma$ considered in this paper does not change over time.
Hence, the following definition of error correction is more precise when the set of attacked sensors does
not change over time.
\begin{definition}
The state $x_k$ is said to be $s$-error correctable after $\vartheta$ steps if there exist a function $F=f(y_k,y_{k+1},\cdots,y_{k+\vartheta-1})$
such that $F=x_k$ for any $x_k\in\mathbb{R}^n$ and any attack sequence $a_{k},$ $a_{k+1},$ $\cdots,a_{k+\vartheta-1}$\vspace{2pt} where \textsc{Supp}$(a_i)\subseteq \Gamma$, $i=k,k+1,\cdots,k+\vartheta-1$ and $\Gamma\subseteq\{1,2,\cdots,q\}$ with
$|\Gamma|= s$.
\label{definition2}
\end{definition}
\par
Using (\ref{syk}) at time step $k-r+1$ and noting that the sensor cannot be attacked when it does not contain any attacked sensor, we get
\begin{align}
y_{k-r+1}(\Gamma)=&C(\Gamma)x_{k-r+1}\label{ykGa}
\end{align}
where $k\geq r-1$.
Similarly, we have
\begin{align}
y_{k-r+2}(\Gamma)=&C(\Gamma)x_{k-r+2}
=C(\Gamma)Ax_{k-r+1}+C(\Gamma)Bu_{k-r+1},\label{ykGb}\\
\vdots&\nonumber\\
y_{k}(\Gamma)=&C(\Gamma)A^{r-1}x_{k-r+1}+\sum_{i=0}^{r-2}C(\Gamma)A^iBu_{k-1-i}\label{ykGc}
\end{align}
where the last equality of (\ref{ykGb}) follows from (\ref{sxk}).
%\begin{align}
%y_{k}(\Gamma)=&C(\Gamma)x_k+a_{k}(\Gamma)\nonumber\\
%=&C(\Gamma)Ax_{k-1}+C(\Gamma)Bu_{k-1}+a_{k}\nonumber\\
%=&C(\Gamma)A^{r-1}x_{k-r+1}+\sum_{i=0}^{r-2}C(\Gamma)A^iBu_{k-1-i}+a_{k}\label{ykGa}
%\end{align}
%where the second equality follows from (\ref{sxk}) and $r\leq n$.
%Similarly, we have
%\begin{align}
%y_{k-1}(\Gamma)=&C(\Gamma)A^{r-2}x_{k-r+1}\!+\!\sum_{i=0}^{r-3}C(\Gamma)A^iBu_{k-2-i}\!+\!a_{k-1},\label{ykGb}\\
%\vdots&\nonumber\\
%y_{k-r+1}(\Gamma)=&C(\Gamma)x_{k-r+1}+a_{k-r+1}(\Gamma).\label{ykGc}
%\end{align}
Putting (\ref{ykGa})$-$(\ref{ykGc}) together, we have
\begin{align}
Y_{k,\Gamma}=O_{\Gamma}x_{k-r+1}+D_{\Gamma}U_{k-1}
\label{ykGd}
\end{align}
where
\begin{align}
&Y_{k,\Gamma}\triangleq\left[
               \begin{array}{c}
                 y_{k-r+1}(\Gamma) \\
                 y_{k-r+2}(\Gamma) \\
                 \vdots\\
                 y_{k}(\Gamma)\\
               \end{array}
             \right],O_{\Gamma}\triangleq\left[
               \begin{array}{c}
                 C(\Gamma) \\
                 C(\Gamma)A \\
                 \vdots\\
                 C(\Gamma)A^{r-1}\\
               \end{array}
             \right],\nonumber\\
&D_{\Gamma}\triangleq\left[
                       \begin{array}{cccc}
                         0 & 0 &  \cdots & 0 \\
                         C(\Gamma)B &0&  \cdots & 0 \\
                         CA(\Gamma)B & C(\Gamma)B & \cdots & 0 \\
                         \vdots & \vdots  & \ddots & \vdots \\
                         C(\Gamma)A^{r-2}B  & C(\Gamma)A^{r-3}B & \cdots & C(\Gamma)B \\
                       \end{array}
                     \right],\nonumber\\
&U_{k-1}\triangleq\left[
               \begin{array}{c}
                 u_{k-r+1} \\
                 u_{k-r+2} \\
                 \vdots\\
                 u_{k-1}\\
               \end{array}
             \right].
\label{ykGe}
\end{align}
\par
Now, we can establish the following lemma, which can exactly solve the state $x_{k-r+1}$ when $\Gamma$ is known.
\begin{lemma} If $\Gamma$ is known and $\big(A,C(\Gamma)\big)$ is observable,
the state $x_{k-r+1}$ can be exactly computed via
\begin{align}
x_{k-r+1}=L_\Gamma(Y_{k,\Gamma}-D_{\Gamma}U_{k-1})
\label{Lem1a}
\end{align}
in $r\geq a$ with $a\leq n$ where
\begin{align}
L_\Gamma\triangleq(O_{\Gamma}^\textrm{T}O_{\Gamma})^{-1}O_{\Gamma}^\textrm{T}\label{Lem1ana}
\end{align}
and $a$ is the minimum value of $r$ such that the rank of $O_{\Gamma}$ is $n$.
 \label{Lemma1}
\end{lemma}
\par
\textit{Proof:}  See Appendix \ref{appensL1}.
\par
\begin{exmp}
Consider the system described by\vspace{1pt} (\ref{sxk}) and (\ref{syk}) where
$A=\left[
     \begin{array}{cc}
       1 & 1 \\
       0 & 1 \\
     \end{array}
   \right]
$, $u_k=0$, $C=\left[
     \begin{array}{cc}
       1 & 2 \\
       1 & 0 \\
       1 & 1 \\
     \end{array}
   \right]$.\vspace{2pt} The first and third sensors can be attacked, that is, $a_{1,k}$ and $a_{3,k}$
can take
arbitrary real number while $a_{2,k}$ is always equal to zero. Then, we have\vspace{2pt} $\Gamma=\{1,3\}$, $y_k(\Gamma)=y_{k,2}$, $C(\Gamma)=\left[
     \begin{array}{cc}
       1 & 0 \\
     \end{array}
   \right]$, $n=2$.
We easily see that $\big(A,C(\Gamma)\big)$ is observable and the rank of $O_{\Gamma}=\left[
     \begin{array}{cc}
       1 & 0 \\
       1 & 1 \\
     \end{array}
   \right]$ is $2$ at $r=2$.
   Then,\vspace{2pt} using (\ref{Lem1ana}), we obtain  $L_{\Gamma}=\left[
     \begin{array}{cc}
       1 & 0 \\
       -1 & 1 \\
     \end{array}
   \right]$.
   Then, using Lemma \ref{Lemma1}, the state $x_{k-1}$ can be computed according to (\ref{Lem1a})\vspace{2pt} where
$Y_{k,\Gamma}=\left[
                \begin{array}{c}
                  y_{k-1,2} \\
                  y_{k,2} \\
                \end{array}
              \right]
$ and $D_{\Gamma}U_{k-1}=0.$ In order to test the correctness of Lemma \ref{Lemma1},\vspace{2pt}  we take \vspace{2pt} $x_{0}=\left[
                \begin{array}{c}
                  2 \\
                  1\\
                \end{array}
              \right]$. Then, we have $Y_{1,\Gamma}=\left[
                \begin{array}{c}
                  2 \\
                  3 \\
                \end{array}
              \right]
$. Then, solving $x_{0}$ via (\ref{Lem1a}) with $k=1$ and $r=2$, we get $x_{0}=\left[
                \begin{array}{c}
                  2 \\
                  1\\
                \end{array}
              \right]$, which means the correctness of Lemma \ref{Lemma1}.
\label{exa1}
\end{exmp}
\par
\begin{remark}
We still take the parameter values of $x_0$, $A$, $u_k$ and $C$ presented in Example \ref{Lemma1} but $\Gamma$ is unknown with $|\Gamma|=2$. Then, noticing $|\Gamma|=2$ and
$q=3$, we have $C_q^{|\Gamma|}=3$ possible cases for set $\Gamma$, namely, $\Gamma=\Gamma_1=\{1,2\}$, $\Gamma=\Gamma_2=\{1,3\}$ or $\Gamma=\Gamma_3=\{2,3\}$.
In  Example \ref{exa1}, we have addressed that $x_{0}$ can be exactly obtained if $\big(A,C(\Gamma)\big)$ is observable at $\Gamma=\Gamma_2$. In the same way,
we conclude that $x_{0}$ can be exactly obtained at $\Gamma=\Gamma_i$ with $i=1,2,3$ if $\big(A,C(\Gamma_i)\big)$ is observable.\vspace{2pt}
Hence, if $\big(A,C(\Gamma_1)\big)$, $\big(A,C(\Gamma_2)\big)$ and $\big(A,C(\Gamma_3)\big)$\vspace{2pt} are all observable, for any $\Gamma=\Gamma_i$, there
exists a $r\leq2$ such that  the state\vspace{2pt} $x_{0}$ is equal to at least an element in the set $\{x_{0}^{(j)}|j=1,2,3\}$ by noticing $x_{0}=x_{0}^{(i)}$ at $\Gamma=\Gamma_i$ obtained from Lemma \ref{Lemma1}
\vspace{2pt}where $x_{0}^{(j)}=L_{\Gamma_j}(Y_{r-1,\Gamma_j}-D_{\Gamma_j}U_{0})$.
In summary, \vspace{2pt}when $\Gamma$ is unknown with $|\Gamma|=2$, we can find a set $\{x_{0}^{(j)}\}$ such that\vspace{2pt} $x_{0}$ is equal to at least an element in the set $\{x_{0}^{(j)}\}$ if the system considered in the remark is 2-sparse
observable
by noting that the following two statements are equivalent obtained from Definition \ref{definition1}:
\begin{itemize}
\item [1.] $\big(A,C(\Gamma_1)\big)$, $\big(A,C(\Gamma_2)\big)$\vspace{2pt} and $\big(A,C(\Gamma_3)\big)$ are all observable.
\item [2.] The system considered in the remark is 2-sparse
observable.
\end{itemize}
As an extension, we will propose a theorem in the following, which addresses that, under proper condition, the state $x_{k-r+1}$ is equal to at least an element in a set when $\Gamma$ is unknown.
\label{Remark1}
\end{remark}
\par
\begin{remark}
\label{Remark2}
It is obvious that there are $C_q^s$ possible cases for $\Gamma$ with $\Gamma\subseteq\{1,2,\cdots,q\}$ and $|\Gamma|=s$. We use $\Gamma_i$ to denote the $i$th possible case
of $\Gamma$. For example, we take $|\Gamma|=2$ and $q=4$. Then, we have $C_4^2$ possible cases for set $\Gamma$, and we can take $\Gamma=\Gamma_1=\{1,2\}$, $\Gamma=\Gamma_2=\{1,3\}$,
$\Gamma=\Gamma_3=\{1,4\}$, $\Gamma=\Gamma_4=\{2,3\}$, $\Gamma=\Gamma_5=\{2,4\}$ or $\Gamma=\Gamma_6=\{3,4\}$ according to common sequential order.
\end{remark}
\begin{theorem}
If the system under consideration is $s$-sparse
observable, there exists a $r$ with $r\leq n$ such that the state $x_{k-r+1}$ is equal to at least an\vspace{2pt} element in the set $X(k,r)=\big\{x_{k-r+1}^{(j)}|j=1,2,\cdots,C_q^s\big\}$\vspace{2pt}
with $x_{k-r+1}^{(j)}=L_{\Gamma_j}(Y_{k,\Gamma_j}-D_{\Gamma_j}U_{k-1})$
where $\Gamma_j$\vspace{2pt} denotes the $j$th possible case
of index set $\Gamma$ defined in Remark \ref{Remark2} with $\Gamma\subseteq\{1,2,\cdots,q\}$ and $|\Gamma|= s$.
\label{Theorem1}
\end{theorem}
\par
\textit{Proof:}  See Appendix \ref{appensT1}.
\par
\begin{exmp}
We take the parameter values of $A$, $u_k$, $C$ and $|\Gamma|$ presented in Remark \ref{Remark1}. Also, we take
$a_{k,1}=2$ and $a_{k,2}=3$ at $\Gamma=\Gamma_1=\{1,2\}$, $a_{k,1}=5.3$ and $a_{k,3}=4$ at $\Gamma=\Gamma_2=\{1,3\}$, and
$a_{k,2}=3$ and $a_{k,3}=4$ at $\Gamma=\Gamma_3=\{2,3\}$.
We easily conclude that the system considered in the example is s-sparse
observable with $s=2$. In order to test\vspace{2pt} the correctness of Theorem \ref{Theorem1}, we take $x_{0}=\left[
                \begin{array}{c}
                  1 \\
                  2\\
                \end{array}
              \right]$,  Noting that $a_{k,3}=0$ at $\Gamma=\Gamma_1$, we see from (\ref{syk}) and (\ref{sxk})
that $y_{0}=\left[\!\!\!
            \begin{array}{ccc}
              7 & 4 & 3 \\
            \end{array}
       \!\!\!   \right]^\textrm{T}
$ and $y_{1}=\left[\!\!\!
            \begin{array}{ccc}
              9 & 6 & 5 \\
            \end{array}
       \!\!\!   \right]^\textrm{T}
$. Then, we see from (\ref{ykGe}) and (\ref{Lem1ana}) that
\begin{align}
&
%              O_{\Gamma_1}=\left[\!
%                                      \begin{array}{cc}
%                                        1 & 1 \\
%                                        1 & 2 \\
%                                      \end{array}
%                                   \! \right],
              L_{\Gamma_1}=\left[\!
                                      \begin{array}{cc}
                                        2 & -1 \\
                                        -1 & 1 \\
                                      \end{array}
                                   \! \right],
%O_{\Gamma_2}=\left[\!
%                    \begin{array}{cc}
%                      1 & 0 \\
%                      1 & 1 \\
%                    \end{array}
%                     \! \right],
%                     O_{\Gamma_3}=\left[\!
%                    \begin{array}{cc}
%                      1 & 2 \\
%                      1 & 3 \\
%                    \end{array}
%                     \! \right],
L_{\Gamma_2}=\left[\!
                    \begin{array}{cc}
                      1 & 0 \\
                      -1 & 1 \\
                    \end{array}
                     \! \right],
L_{\Gamma_3}=\left[\!
                    \begin{array}{cc}
                      3 & -2 \\
                      -1 & 1 \\
                    \end{array}
                     \! \right],
                                                  \nonumber\\
&Y_{1,\Gamma_1}=\left[\!
                 \begin{array}{c}
                   3 \\
                   5\\
                 \end{array}
               \!\right],Y_{1,\Gamma_2}=\left[\!
                 \begin{array}{c}
                   4 \\
                   6\\
                 \end{array}
              \! \right], Y_{1,\Gamma_3}=\left[\!
                 \begin{array}{c}
                   7 \\
                   9\\
                 \end{array}
               \!\right],
                  D_{\Gamma_j}U_{0}=\left[\!
                 \begin{array}{c}
                   0 \\
                   0\\
                 \end{array}
               \!\right]
\nonumber
\end{align}
with $j=1,2,3$ when we select $r=2$. Then, we see from Theorem \ref{Theorem1} that
\begin{align}
x_{0}^{(1)}=\left[\!
                 \begin{array}{c}
                   1 \\
                   2\\
                 \end{array}
               \!\right], x_{0}^{(2)}=\left[\!
                 \begin{array}{c}
                   4 \\
                   2\\
                 \end{array}
               \!\right], x_{0}^{(3)}=\left[\!
                 \begin{array}{c}
                   3 \\
                   2\\
                 \end{array}
               \!\right]                              \nonumber
\end{align}
where $x_{0}=x_{0}^{(1)}$. Hence, the solution of $x_{0}$ is equal\vspace{2pt} to at least an element $x_{0}^{(1)}$ in the set $X(1,2)=\big\{x_{0}^{(j)}|j=1,2,3\big\}$.
In the same way, we have \begin{align}
x_{0}^{(1)}=\left[\!
                 \begin{array}{c}
                   5 \\
                   2\\
                 \end{array}
               \!\right], x_{0}^{(2)}=\left[\!
                 \begin{array}{c}
                   1 \\
                   2\\
                 \end{array}
               \!\right], x_{0}^{(3)}=\left[\!
                 \begin{array}{c}
                   6.3 \\
                   2\\
                 \end{array}
               \!\right]                              \nonumber
\end{align} at $\Gamma=\Gamma_2$ and
\begin{align}
x_{0}^{(1)}=\left[\!
                 \begin{array}{c}
                   5 \\
                   2\\
                 \end{array}
               \!\right], x_{0}^{(2)}=\left[\!
                 \begin{array}{c}
                   4 \\
                   2\\
                 \end{array}
               \!\right], x_{0}^{(3)}=\left[\!
                 \begin{array}{c}
                   1 \\
                   2\\
                 \end{array}
               \!\right]                              \nonumber
\end{align} at $\Gamma=\Gamma_3$.
Hence, we have $x_{0}=\left[
                \begin{array}{c}
                  1 \\
                  2\\
                \end{array}
              \right]$\vspace{2pt} is equal to at least an element in $X(1,2)$ no matter which possible case $\Gamma$ takes, which identifies the correctness of
              Theorem \ref{Theorem1}.
\label{exa2}
\end{exmp}
\par
\begin{remark}
We see from Theorem \ref{Theorem1} that there exists a $r$ with $r\leq n$ such that \vspace{2pt} $x_{k-r+1}$ is equal to at least an element in the set $X(k,r)$ if the system under consideration is $s$-sparse
observable, which is
identified in Example \ref{exa2}. This means that the state is $s$-error correctable after $r$ steps with $r\leq n$ if the system under consideration is $s$-sparse
observable. In \cite{Sr1}, it was addressed that the state is $s$-error correctable if $|\textrm{supp}(Cz)\cup\cdots\cup\textrm{supp}(CA^{r-1}z)|>2s$ (considering the notation definition in this paper) for all
$z\in \mathbb{R}^n\backslash \{0\}$ and that
the maximum number of attacked sensors $s$ must be less than $\lceil q/2\rceil$ so that
 the state is $s$-error correctable.
Similar results presented in \cite{Sr2} addressed that the state is $s$-error correctable if the system is $2s$-sparse
observable, which means that the state is not correctable
if there are $\lceil q/2\rceil$ or more sensors are attacked. However, from Theorem \ref{Theorem1} and Example \ref{exa2}, we find that when the system is $s$-sparse
observable, the state is correctable
even though there are $\lceil q/2\rceil$ or more sensors are attacked. Hence, compared with Theorem \ref{Theorem1},
the conditions of error correction presented in \cite{Sr1} and \cite{Sr2} are more conservative.\vspace{2pt} When the set of attacked sensors
does\vspace{2pt} not change, we have $|a_k^{(a)}\bigcup a_k^{(b)}|\leq s$ where $a_k^{(a)}$ and $a_k^{(b)}$ denote two different values of $a_k$.\vspace{2pt}
The sets of attacked sensors considered in \cite{Sr1} and \cite{Sr2}
do not change but $|a_k^{(a)}\bigcup a_k^{(b)}|\leq 2s$ (considering the notation definition in this paper) instead of $|a_k^{(a)}\bigcup a_k^{(b)}|\leq s$  was used, which leads to more conservative results.
\label{Remark3}
\end{remark}
\begin{remark}
From the proof of Theorem \ref{Theorem1}, we easily see that, under the condition of  $s$-sparse observability, the state cannot be corrected in $r<b$  where $b$ is defined in Lemma \ref{LeThe1}. This is because
we cannot obtain rank$(O_{\Gamma_j})=n$\vspace{2pt} for any $j$ with $j=1,2,\cdots,C_q^s$ when  $r<b$. Hence, for the error correction result presented in this paper, the minimum number of steps such that the state is $s$-error correctable is $r=b$.
In this paper, $b$ is referred to as the $s$-sparse observable lower bound.
\label{Remark3k}
\end{remark}
\subsection{Secure State Reconstruction based on SESVS \label{secmrb}}
\begin{remark}
Although the state is $s$-error correctable under the condition of  $s$-sparse observability presented in Theorem \ref{Theorem1} and Remark \ref{Remark3}, it is impossible to find the state among a set if only one element
in the set is equal to the state.  If there exists a set such that more than one element in the set is equal to the state $x_{k-r+1}$, we can
find the state by searching elements with the same value in the set. We will provide the following theorem to solve this problem.
\label{Remark4}
\end{remark}
\par
\begin{theorem}
If the system under consideration is $(s+1)$-sparse
observable satisfying $q\geq s+2$, there exists a $r$ with $r\leq n$ such that the\vspace{2pt} state $x_{k-r+1}$ is equal to at least $q-s$ elements in the set $Y(k,r)=\big\{y_{k-r+1}^{(j)}|j=1,2,\cdots,C_q^{s+1}\big\}$\vspace{2pt}
with $y_{k-r+1}^{(j)}=L_{\Lambda_j}(Y_{k,\Lambda_j}-D_{\Lambda_j}U_{k-1})$
where $\Lambda_j$\vspace{2pt} denotes the $j$th possible case
of index set $\Lambda$ with $\Lambda\subseteq\{1,2,\cdots,q\}$ and $|\Lambda|=s+1$.
\label{Theorem2}
\end{theorem}
\par
\textit{Proof:}  See Appendix \ref{appenT21}.
\par
\begin{exmp}
We still take the parameter values of $x_0$, $A$, $u_k$ and $C$ presented in Example \ref{exa1}, and $\Gamma$ is unknown with $|\Gamma|=1$.
Hence, we have three possible cases for $\Gamma$, namely, $\Gamma=\Gamma_1=\{1\}$, $\Gamma=\Gamma_2=\{2\}$ and $\Gamma=\Gamma_3=\{3\}$.
We take $a_{k,1}=3.5$ in $\Gamma=\Gamma_1$, $a_{k,2}=2$ in $\Gamma=\Gamma_2$ and $a_{k,3}=1$ in $\Gamma=\Gamma_3$.
We easily see that the system considered in the remark is $2$-sparse
observable with $q= s+2$, $\Lambda_1=\{1,2\}$, $\Lambda_2=\{1,3\}$ and $\Lambda_3=\{2,3\}$.
Then, after taking $r=2$, we see from Theorem \ref{Theorem2} that
$Y(1,2)=\big\{y_{0}^{(j)}|j=1,2,3\big\}$ with
\begin{align}
y_{0}^{(j)}=L_{\Lambda_j}(Y_{1,\Lambda_j}-D_{\Lambda_j}U_{0})
\label{exam3a}
\end{align}
where $L_{\Lambda_j}$, $Y_{1,\Lambda_j}$, $D_{\Lambda_j}$ and $U_{0}$ are obtained using (\ref{ykGe}) and (\ref{Lem1ana}).
Noting that $a_{k,2}=a_{k,3}=0$ in $\Gamma=\Gamma_1$, we see from (\ref{syk}) and (\ref{sxk})
that $y_{0}=\left[\!\!\!
            \begin{array}{ccc}
              7.5 & 2 & 3 \\
            \end{array}
       \!\!\!   \right]^\textrm{T}
$ and $y_{1}=\left[\!\!\!
            \begin{array}{ccc}
              8.5 & 3 & 4 \\
            \end{array}
       \!\!\!   \right]^\textrm{T}
$.
Using (\ref{exam3a}) at $\Gamma=\Gamma_1$, we have
\begin{align}
y_{0}^{(1)}=\left[\!
                 \begin{array}{c}
                   2 \\
                   1\\
                 \end{array}
               \!\right], y_{0}^{(2)}=\left[\!
                 \begin{array}{c}
                   2 \\
                   1\\
                 \end{array}
               \!\right], y_{0}^{(3)}=\left[\!
                 \begin{array}{c}
                   5.5 \\
                   1\\
                 \end{array}
               \!\right]
\nonumber
\end{align}
where $x_0=y_{0}^{(1)}=y_{0}^{(2)}$.
Similarly,  we have
\begin{align}
y_{0}^{(1)}=\left[\!
                 \begin{array}{c}
                   2 \\
                   1\\
                 \end{array}
               \!\right], y_{0}^{(2)}=\left[\!
                 \begin{array}{c}
                   4 \\
                   1\\
                 \end{array}
               \!\right], y_{0}^{(3)}=\left[\!
                 \begin{array}{c}
                   2 \\
                   1\\
                 \end{array}
               \!\right]
\nonumber
\end{align} at $\Gamma=\Gamma_2$
and
\begin{align}
y_{0}^{(1)}=\left[\!
                 \begin{array}{c}
                   3 \\
                   1\\
                 \end{array}
               \!\right], y_{0}^{(2)}=\left[\!
                 \begin{array}{c}
                   2 \\
                   1\\
                 \end{array}
               \!\right], y_{0}^{(3)}=\left[\!
                 \begin{array}{c}
                   2 \\
                   1\\
                 \end{array}
               \!\right]
\nonumber
\end{align} at $\Gamma=\Gamma_3$.
Hence, the state $x_0$ is equal to at least $2$ elements in the set $Y(1,2)$ no matter which possible case $\Gamma$ takes. This indicates the correctness of Theorem \ref{Theorem2}.
\label{exa3}
\end{exmp}
\par
\par
The following corollary provides an extension of Theorem \ref{Theorem2}.
\par
\begin{corollary}
If the system under consideration is $(s+\tau)$-sparse
observable satisfying $q\geq s+\tau+1$ and $\tau\geq1$, there exists a $r$ with $r\leq n$ such that the\vspace{2pt} state $x_{k-r+1}$ is equal to at least
$C_{q-s}^\tau$ elements in the set $Y_\tau(k,r)=\big\{y_{k-r+1}^{(j)}|j=1,2,\cdots,C_q^{s+\tau}\big\}$\vspace{2pt}
with $y_{k-r+1}^{(j)}=L_{\Lambda_j}(Y_{k,\Lambda_j}-D_{\Lambda_j}U_{k-1})$
where $\Lambda_j$\vspace{2pt} denotes the $j$th possible case
of index set $\Lambda^{[\tau]}$ with $\Lambda^{[\tau]}\subseteq\{1,2,\cdots,q\}$ and $|\Lambda^{[\tau]}|=s+\tau$.
\label{corollary1}
\end{corollary}
\par
\textit{Proof:}  See Appendix \ref{appenT22}.
\par
\begin{remark}
From Theorem \ref{Theorem2}, we see that the state $x_{k-r+1}$ can be accurately reconstructed unless the adversary can create at least $q-s$ elements with the same value in the set $Y(k,r)$
such that the value is not equal to the state $x_{k-r+1}$. In this case, there are two subsets of $Y(k,r)$ where all elements in either subset have the same value and
the element's value in a subset is not equal to the element's value in the other subset. As a result, we do not know which element is equal to the real state.  In the following theorem,
we address the conditions the adversary must obtain so that the state $x_{k-r+1}$ cannot be reconstructed via Theorem \ref{Theorem2}.
\label{Remark6}
\end{remark}
\begin{theorem}
When the system under consideration is $(s+1)$-sparse
observable satisfying $q\geq s+2$, there are at least $q-s$ elements in the set $Y(k,r)$ such that the elements have the same value that is not equal to the state $x_{k-r+1}$
if and only if there exists $\Upsilon_{1}$, $\Upsilon_{2}$, $\cdots$,
$\Upsilon_{q-s}$ such that
\begin{align}
&L_{\Upsilon_1}A_{k,\Upsilon_1}
=L_{\Upsilon_\rho}A_{k,\Upsilon_\rho},\label{the3a2}\\
&L_{\Upsilon_1}A_{k,\Upsilon_1}\neq0\label{the3a1}
\end{align}
for any $\rho$, $\rho=1,2,\cdots,q-s$, where $\Upsilon_{1}$, $\Upsilon_{2}$, $\cdots$,
$\Upsilon_{q-s}$ denote $q-s$ different cases of index set $\Lambda$ defined in Theorem \ref{Theorem2} and $A_{k,\Upsilon_\rho}$ is defined as
\begin{align}
A_{k,\Upsilon_\rho}\triangleq \left[
               \begin{array}{c}
                 a_{k-r+1}(\Upsilon_\rho) \\
                 a_{k-r+2}(\Upsilon_\rho) \\
                 \vdots\\
                 a_k(\Upsilon_\rho)\\
               \end{array}
             \right].\label{the3a3}
\end{align}
\label{Theorem3}
\end{theorem}
\par
\textit{Proof:}  See Appendix \ref{appenT3}.
\par
Making reference to the proof of Theorem \ref{Theorem3}, we easily obtain the following corollary which is an extension of Theorem  \ref{Theorem3}.
\par
\begin{corollary}
When the system under consideration is $(s+\tau)$-sparse
observable satisfying $q\geq s+\tau+1$, there are at least $C_{q-s}^\tau$ elements in the set $Y_\tau(k,r)$ such that the elements have the same value that is not equal to the state $x_{k-r+1}$
if and only if there exists $\Upsilon_{1}$, $\Upsilon_{2}$, $\cdots$,
$\Upsilon_{C_{q-s}^\tau}$ such that
\begin{align}
&L_{\Upsilon_1}A_{k,\Upsilon_1}
=L_{\Upsilon_\rho}A_{k,\Upsilon_\rho},\nonumber\\
&L_{\Upsilon_1}A_{k,\Upsilon_1}\neq0\nonumber
\end{align}
for any $\rho$, $\rho=1,2,\cdots,C_{q-s}^\tau$, where $\Upsilon_{1}$, $\Upsilon_{2}$, $\cdots$,
$\Upsilon_{C_{q-s}^\tau}$ denote $C_{q-s}^\tau$ different cases of index set $\Lambda^{[\tau]}$ defined in Corollary \ref{corollary1}.
\label{corollary2}
\end{corollary}
\par
\begin{remark}
From Theorem  \ref{Theorem3},\vspace{2pt} we see that, in order to get at least $q-s$ elements $z_{k-r+1}^{(1)}$, $z_{k-r+1}^{(2)}$, $\cdots$, $z_{k-r+1}^{(q-s)}$\vspace{2pt} in the set $Y(k,r)$ such that $z_{k-r+1}^{(1)}=z_{k-r+1}^{(2)}=\cdots=z_{k-r+1}^{(q-s)}\neq x_{k-r+1}$,\vspace{2pt} the adversary must control the value of $A_{\Upsilon_\rho}$ so that
the conditions given in (\ref{the3a2}) and (\ref{the3a1}) are simultaneously satisfied.
Hence, to prevent the state reconstruction using Theorem \ref{Theorem2}, the following requirements must be simultaneously satisfied:
\begin{itemize}
  \item The adversary can exactly\vspace{2pt} knows $L_{\Upsilon_{1}}$, $L_{\Upsilon_{2}}$, $\cdots$, $L_{\Upsilon_{q-s}}$.
  \item  The solution for
$L_{\Upsilon_1}A_{k,\Upsilon_1}
=L_{\Upsilon_\rho}A_{k,\Upsilon_\rho}$ with $L_{\Upsilon_1}A_{k,\Upsilon_1}\neq0$ and $\rho=1,2,\cdots,q-s$\vspace{2pt} exists.
  \item The adversary\vspace{2pt} can solve $A_{k,\Upsilon_\rho}$ through $L_{\Upsilon_1}A_{k,\Upsilon_1}
=L_{\Upsilon_\rho}A_{k,\Upsilon_\rho}$ with $L_{\Upsilon_1}A_{k,\Upsilon_1}\neq0$.\vspace{2pt}
  \item The adversary can attack the corresponding sensors
using $a_{\varsigma,\Upsilon_\rho}$ at time step $\varsigma$ with $\varsigma=k-r+1,k-r+2,\cdots,k$.
\end{itemize}
So, it is very hard to prevent the state reconstruction when we use Theorem \ref{Theorem2} for reconstructing the state.
Even if the adversary can prevent the state reconstruction when the above requirements are simultaneously satisfied, we still can reconstruct the state
using Corollary \ref{corollary1} under proper condition.
More precisely, when we cannot reconstruct the state using Theorem \ref{Theorem2},
in order to reconstruct the state, we can use Corollary \ref{corollary1} for yielding the set $Y_\tau(k,r)$ under the condition of $(s+\tau)$-sparse observability such that
there are at least $C_{q-s}^\tau$ elements with the same value in  $Y_\tau(k,r)$.
In this case, we can find the real state from the set $Y_\tau(k,r)$ unless the adversary can create at least $C_{q-s}^\tau$ elements in $Y_\tau(k,r)$ such that the $C_{q-s}^\tau$ elements have the same value and are not equal to the real state.
When the adversary cannot create at least $C_{q-s}^\tau$\vspace{2pt} elements satisfying the above mentioned requirement in $Y_\tau(k,r)$, that is, the conditions presented in
Corollary \ref{corollary2} cannot be satisfied, we can definitely reconstruct the state from $Y_\tau(k,r)$.
\label{Remark7}
\end{remark}
\begin{remark}
It is worth mentioning that we can definitely reconstruct the state $x_{k-r+1}$ from the set $Y_\tau(k,r)$ under the condition of $(s+\tau)$-sparse observability if
$C_q^{s+\tau}<2C_{q-s}^\tau$. Using Corollary \ref{corollary1}, we see that there exists at least $C_{q-s}^\tau$ elements in the set $Y_\tau(k,r)$ such that
the $C_{q-s}^\tau$ elements have the same value and are equal to the state $x_{k-r+1}$ if the condition of $(s+\tau)$-sparse observability is satisfied.
It is obvious that the number of the remaining elements in $Y_\tau(k,r)$ is $C_q^{s+\tau}-C_{q-s}^\tau$\vspace{2pt}
and is less than $C_{q-s}^\tau$ if $C_q^{s+\tau}<2C_{q-s}^\tau$. Hence,\vspace{1pt} if $C_q^{s+\tau}<2C_{q-s}^\tau$, we\vspace{1pt} can definitely find the real state from searching
$C_{q-s}^\tau$ elements with the same value in  $Y_\tau(k,r)$ since the number of the remaining elements in $Y_\tau(k,r)$ is less than $C_{q-s}^\tau$. \vspace{2pt}The correctness of the statement is verified by Example
\ref{exa3}
with $q=3$, $\tau=1$ and $s=1$.
\label{Remark8}
\end{remark}
\vspace{-5pt}
\subsection{Secure State Reconstruction based on SESGC \label{secmrc}}
\begin{lemma}
If the system under consideration is $s$-sparse
observable,
there\vspace{2pt} exist at least a $r$ with $r\leq n$ and a $\mu$ such that  $\mu \in \{1,2,\cdots,C_q^s\}$,
$x_{k-r+1}^{(\mu)}=x_{k-r+1}$\vspace{2pt} and $x_{k-r+2}^{(\mu)}-Ax_{k-r+1}^{(\mu)}-Bu_{k-r+1}=0$.\vspace{2pt}
 \label{LeSPEa}
\end{lemma}
\par
\textit{Proof:} Using Theorem \ref{Theorem1}, \vspace{2pt} we see that there exist at least a $r$ with $r\leq n$ and a $\mu$
such that $\mu \in \{1,2,\cdots,C_q^s\}$, $x_{k-r+1}^{(\mu)}=x_{k-r+1}$,\vspace{2pt} and $x_{k-r+2}^{(\mu)}=x_{k-r+2}$.
Hence, next, we only need to prove $x_{k-r+2}^{(\mu)}-Ax_{k-r+1}^{(\mu)}-Bu_{k-r+1}=0$.
From (\ref{sxk}), it follows that
\begin{align}
x_{k-r+2}-Ax_{k-r+1}-Bu_{k-r+1}=0.\label{PLeSPEa1}
\end{align}
Substituting\vspace{2pt} $x_{k-r+1}^{(\mu)}=x_{k-r+1}$ and $x_{k-r+2}^{(\mu)}=x_{k-r+2}$ into (\ref{PLeSPEa1}), we obtain $x_{k-r+2}^{(\mu)}-Ax_{k-r+1}^{(\mu)}-Bu_{k-r+1}=0$.\vspace{2pt}
This completes the proof of Lemma \ref{LeSPEa}.
\par
\begin{remark}
From Lemma \ref{LeSPEa}, we see that there\vspace{2pt} exist at least a $x_{k-r+1}^{(\mu)}\in X(k,r)$ and a $x_{k-r+2}^{(\mu)}\in X(k+1,r)$\vspace{2pt} such that
$x_{k-r+1}^{(\mu)}=x_{k-r+1}$\vspace{2pt} and $x_{k-r+2}^{(\mu)}-Ax_{k-r+1}^{(\mu)}-Bu_{k-r+1}=0.$  Hence, we can reconstruct
the state $x_{k-r+1}$ that is equal\vspace{2pt} to any element in $\check{X}_1(k,r)$ if\vspace{2pt} there only exists an element in $\check{X}_1(k,r)$ or each element in $\check{X}_1(k,r)$\vspace{2pt} has the same value
where the set $\check{X}_1(k,r)$ is obtained from searching $x_{k-r+1}^{(j)}\in X(k,r)$\vspace{2pt} from $j=1$ to $j=C_q^s$
such that $x_{k-r+2}^{(j)}-Ax_{k-r+1}^{(j)}-Bu_{k-r+1}=0.$\vspace{2pt}
Inspired by the above strategy of searching element $x_{k-r+1}^{(j)}$\vspace{2pt} satisfying the condition $x_{k-r+2}^{(j)}-Ax_{k-r+1}^{(j)}-Bu_{k-r+1}=0$,\vspace{2pt} we propose the following algorithm to reconstruct the state.
\label{Remark5B}
\end{remark}
\begin{tabular*}{8.86cm}{c}
  \thickhline
\end{tabular*}
\\
 \textbf{Algorithm 1:}\! \textbf{Secure State Reconstruction based on SESGC}
\begin{tabular*}{8.86cm}{c}
  \hline
\end{tabular*}
\\
\textbf{Step 1:} Obtain the sets $D_1$ and $\check{X}_1(k,r)$.\par
 \ \  $D_1$ is obtained via the following scheme:\vspace{3pt}
\par
\hspace{15pt}\textbf{for}\vspace{3pt} $j=1:C_q^s$\par
 \hspace{20pt} \textbf{if} \hspace{5pt}
 $x_{k-r+2}^{(j)}-Ax_{k-r+1}^{(j)}-Bu_{k-r+1}=0$, \textbf{then} $j\in D_1$\par\vspace{3pt}
\hspace{20pt}  \textbf{else}\ \  $j\notin D_1$\par\vspace{3pt}
\hspace{20pt} \textbf{end if} \par\vspace{3pt}
\hspace{12.5pt} \textbf{end for} \vspace{3pt}
\\
where $x_{k-r+\varsigma}^{(j)}$ with \vspace{1pt} $\varsigma=1,2$ denotes the $j$th element of $X(k+\varsigma-1,r)$ and $x_{k-r+\varsigma}^{(j)}$ is computed according to\vspace{2pt}
$x_{k-r+\varsigma}^{(j)}=L_{\Gamma_j}(Y_{k+\varsigma-1,\Gamma_j}-D_{\Gamma_j}U_{k+\varsigma-2})$\vspace{1pt} where $r$ is equal to the $s$-sparse observable lower bound, namely,
$r=b$ with $b=\underset{_{i\in \{1,2,\cdots,C_q^s\}}}{\textrm{max}}r_i$
where $r_i$ is the minimum value of $r$ such that the rank of $O_{\Gamma_i}$ is $n$. Then, $\check{X}_1(k,r)$\vspace{2pt} is obtained according to
$\check{X}_1(k,r)=\big\{x_{k-r+1}^{(j)}|j\in D_1\big\}$. \vspace{2pt}
\\
\textbf{Step 2:} If there only exists an element in $\check{X}_1(k,r)$ or each element in $\check{X}_1(k,r)$ has the same value, we can obtain the state $x_{k-r+1}$\vspace{2pt} which is equal to any element in the set
 $\check{X}_1(k,r)$.
\\
\textbf{Step 3:} Otherwise, that is, \vspace{2pt}
all elements in the set $\check{X}_1(k,r)$ do not have the same value,\vspace{2pt} we obtain the sets\vspace{2pt} $D_2$ and $\check{X}_2(k,r)$  where
$D_2$ is obtained via the following scheme:\vspace{3pt}
\par
\hspace{15pt}\textbf{for}\vspace{3pt} $\forall j\in D_1$\par
 \hspace{20pt} \textbf{if} \hspace{5pt}
 $x_{k-r+3}^{(j)}-Ax_{k-r+2}^{(j)}-Bu_{k-r+2}=0$, \textbf{then} $j\in D_2$\par\vspace{3pt}
\hspace{20pt}  \textbf{else}\ \  $j\notin D_2$\par\vspace{3pt}
\hspace{20pt}  \textbf{end if}\par\vspace{3pt}
\hspace{12.5pt} \textbf{end for} \vspace{3pt}\\
Then, $\check{X}_2(k,r)$\vspace{2pt} is obtained according to
$\check{X}_2(k,r)=\big\{x_{k-r+1}^{(j)}|j\in D_2\big\}$. \vspace{2pt}
\\
\textbf{Step 4:} If there only exists an element in $\check{X}_2(k,r)$ or each element in $\check{X}_2(k,r)$ has the same value, we can obtain the state $x_{k-r+1}$\vspace{2pt} which is equal to any element in the set
 $\check{X}_2(k,r)$. Otherwise, we obtain $\check{X}_3(k,r)$, $\check{X}_4(k,r),\ \cdots$, $\check{X}_\vartheta(k,r)$ until
there only exists an element in $\check{X}_\vartheta(k,r)$ or each element in $\check{X}_\vartheta(k,r)$ has the same value
where $\check{X}_\varsigma(k,r)$ with $\varsigma=3,4,\cdots,\vartheta$ is recursively computed according to the following scheme:
\vspace{3pt}
%\par
%\hspace{15pt}for\vspace{3pt}  $i=2:\varsigma$\par
%\vspace{3pt}
\par
\hspace{15pt}\textbf{for}\vspace{3pt} $\forall j\in D_{\varsigma-1}$\par
 \hspace{20pt} \textbf{if} \hspace{5pt}
 $x_{k-r+\varsigma+1}^{(j)}-Ax_{k-r+\varsigma}^{(j)}-Bu_{k-r+\varsigma}=0$, \textbf{then} $j\in D_\varsigma$\par\vspace{3pt}
\hspace{20pt}  \textbf{else}\ \  $j\notin D_\varsigma$\par\vspace{3pt}
\hspace{20pt}  \textbf{end if}\par\vspace{3pt}
\hspace{12.5pt} \textbf{end for} \vspace{3pt}\par
\hspace{12.5pt} Then, $\check{X}_\varsigma(k,r)=\big\{x_{k-r+1}^{(j)}|j\in D_\varsigma\big\}$.\\
\begin{tabular*}{8.86cm}{c}
  \thickhline
\end{tabular*}
\par
The proof of Algorithm 1 is given in Appendix \ref{appenA1a}.
\begin{remark}
Using Algorithm 1, we can accurately reconstruct the state $x_{k-r+1}$ unless the adversary can create\vspace{2pt} at least a $x_{k-r+1}^{(v)}$ such that\vspace{2pt}
$v\in D_{\varsigma}$, $\varsigma=1,2,\cdots,$ $x_{k-r+1}^{(v)}\neq x_{k-r+1}$ and $x_{k-r+\varsigma+1}^{(v)}-Ax_{k-r+\varsigma}^{(v)}-Bu_{k-r+\varsigma}=0$.\vspace{2pt}
In this case, there exists at least two elements with different values in the set  $\check{X}_\varsigma(k,r)$. As a result, we do not know which element in the set is equal to the
real state. In the following theorem, we address the conditions the adversary must obtain so that the state $x_{k-r+1}$ cannot be reconstructed using Algorithm 1.
\label{Remark5b}
\end{remark}
\begin{theorem} When the system under consideration is $s$-sparse
observable,
there exists at least an element in the set $\check{X}_\varsigma(k,r)$ such that the state $x_{k-r+1}$ is not equal to the element
if and only if there\vspace{2pt} exists at least a $v$ with $v\in D_{\varsigma}$ such that $L_{\Gamma_v}A_{k,\Gamma_v}\neq0$ and $L_{\Gamma_v}A_{k+\rho,\Gamma_v}-AL_{\Gamma_v}A_{k+\rho-1,\Gamma_v}=0$,
$\rho=1,2,\cdots,\varsigma$.
\label{Theorem4}
\end{theorem}
\par
\textit{Proof:}  See Appendix \ref{appenA1b}.
\par
\begin{remark}
When Algorithm 1 is used to reconstruct the state,
 we see from Theorem \ref{Theorem4} that the adversary has the chance to prevent the state reconstruction if the adversary can exactly know $L_{\Gamma_v}$ and the solution for
$L_{\Gamma_v}A_{k+\rho,\Gamma_v}-AL_{\Gamma_v}A_{k+\rho-1,\Gamma_v}=0$ with $L_{\Gamma_v}A_{k,\Gamma_v}\neq0$ and $\rho=1,2,\cdots$ exists.
However, it is very hard for the adversary
to prevent the state reconstruction even though the adversary can\vspace{2pt} exactly know $L_{\Gamma_v}$ and the solution for
$L_{\Gamma_v}A_{k+\rho,\Gamma_v}-AL_{\Gamma_v}A_{k+\rho-1,\Gamma_v}=0$ exists.\vspace{2pt}
To prevent the state reconstruction, the following requirements must be simultaneously satisfied:
\begin{itemize}
  \item The adversary can exactly\vspace{2pt} knows $L_{\Gamma_v}$.
  \item  The solution for
$L_{\Gamma_v}A_{k+\rho,\Gamma_v}-AL_{\Gamma_v}A_{k+\rho-1,\Gamma_v}=0$ with $L_{\Gamma_v}A_{k,\Gamma_v}\neq0$ and $\rho=1,2,\cdots$\vspace{2pt} exists.
  \item The adversary\vspace{2pt} can solve $A_{k+\rho,\Gamma_v}$ via $L_{\Gamma_v}A_{k+\rho,\Gamma_v}-\vspace{2pt}AL_{\Gamma_v}A_{k+\rho-1,\Gamma_v}=0$
  with $L_{\Gamma_v}A_{k,\Gamma_v}\neq0$.
  \item The adversary can attack the corresponding sensors
using $a_{\varsigma,\Gamma_v}$ at time step $\varsigma$ with $\varsigma=k-r+1,k-r+2,\cdots,k+\rho$.
\end{itemize}
Hence, it is very hard
to prevent the state reconstruction when Algorithm 1 is used to reconstruct the state.
\label{Remark10}
\end{remark}
\par
\begin{remark}
In order to reconstruct the state from measurements with attacked sensors,
a lot of methods were proposed.
The authors tried to reconstruct the state by solving optimization problem in \cite{Sr1} and then
some methods for reconstructing the state based on solving optimization problem were proposed \cite{Sr2}$-$\cite{Sr4}, \cite{Sr6}, \cite{Sr7}, \cite{Srm3}.
An event-triggered projected Luenberger observer was proposed in \cite{Sr2} to reconstruct the state and more results for secure state reconstruction based on designing observers were proposed in
\cite{Sr4}, \cite{Sr5} and \cite{Sr8}.
Different from the previously proposed methods, the first method proposed in Section \ref{secmrb} is based  on the idea of SESVS and the second method proposed in Algorithm 1 is
based on SESGC. \vspace{1pt}
More precisely, Algorithm 1 is obtained from
 searching each element in the set $\check{X}_i(k,r)$ from $i=1$ to $i=\vartheta$ such that:
\begin{itemize}
  \item There only exists an element in $\check{X}_\vartheta(k,r)$ or each element in $\check{X}_\vartheta(k,r)$ has the same value.
  \item The elements in the set $\check{X}_j(k,r)$ with $j=1,2,\cdots,\vartheta-1$ are not all equal
\end{itemize}\vspace{2pt}
where $\check{X}_i(k,r)=\big\{x_{k-r+1}^{(j)}|j\in D_i\big\}$\vspace{2pt} and $D_i$ is obtained from searching\vspace{2pt} element $j$ satisfying
$x_{k-r+i+1}^{(j)}-Ax_{k-r+i}^{(j)}-Bu_{k-r+i}=0$ and $j\in D_{i-1}$ with $D_0=\big\{1,2,\cdots,C_q^s\big\}$.
Compared with the previous results of secure state reconstruction, the two proposed methods have the following two advantages:
\begin{itemize}
  \item The two proposed methods still can reconstruct the state in $s\geq\lceil q/2\rceil$ under proper conditions while the previous results cannot reconstruct the state in $s\geq\lceil q/2\rceil$.
  \item Using either of the two proposed methods, we can accurately reconstruct the state while the reconstructed values
can only gradually approach the real state using the previous methods which were obtained from solving optimization problems or designing observers.
\end{itemize}
\label{Remark5}
\end{remark}
\begin{remark}
A comparison of the two proposed methods is given as follows:
\begin{itemize}
  \item The state reconstruction method based on SESVS, including the results proposed in Theorem \ref{Theorem2} and Corollary \ref{corollary1}, requires the condition of $(s+\tau)$-sparse observability with $1\leq\tau\leq q-s-1$ and the measurements from time steps $k-r+1$ to $k$ where $r$ is determine via the corresponding $(s+\tau)$-sparse observable lower bound.
  \item The state reconstruction method based on SESGC, that is, Algorithm 1, requires the condition of $s$-sparse observability and the measurements from time steps $k-r+1$ to $k+\vartheta$ with $\vartheta\geq 1$
   where $r$ is determined via the corresponding $s$-sparse observable lower bound.
   \item For the computational complexity, the state reconstruction method based on SESVS requires computing $C_q^{s+\tau}$ elements, the method based on SESGC requires computing at least $2C_q^{s}$ elements.
\end{itemize}
Hence, for the two proposed methods, the method based on SESVS usually requires less computational complexity and the method based on SESGC requires a less conservative condition.
\label{Remark11}
\end{remark}
\begin{remark}
From Theorem \ref{Theorem1} and the corresponding proof, we see that there exists a positive integer $b$ with $b=\underset{_{i\in \{1,2,\cdots,C_q^s\}}}{\textrm{max}}r_i\leq n$ such that the state $x_{k-r+1}$ with
$r=b$ is equal to at least an\vspace{2pt} element in the set $X(k,r)$ if the system under consideration is $s$-sparse
observable.
 We easily conclude that the statement holds in $r\geq b$, that is, there exists a positive integer $b$ with $b\leq n$ such that the state $x_{k-r+1}$ is equal to at least an\vspace{2pt} element in the set $X(k,r)$ in $r\geq b$ if the system under consideration is $s$-sparse
observable. Similarly, for Theorem \ref{Theorem2}, Corollary \ref{corollary1} and Algorithm 1, we can easily obtain the corresponding extensions.
For example, as an extension of Theorem \ref{Theorem2}, the following statement holds:
\\
If the system under consideration is $(s+1)$-sparse
observable satisfying $q\geq s+2$, there exists a positive
integer $b_1$ with $b_1\leq n$ such that the\vspace{2pt} state $x_{k-r+1}$ is equal to at least $q-s$ elements in the set $Y(k,r)$ when $r\geq b_1$ where $b_1$ is the $(s+1)$-sparse observable lower bound.
\\
For the real-world example of three-inertia systems provided in Section \ref{sectionNEb}, the condition of Theorem \ref{Theorem2} is satisfied at $r=3$.
However, due to the singular
matrix error in running MATLAB, the state cannot be reconstructed via Theorem \ref{Theorem2} when we take $r=3$. Hence, in the example of three-inertia systems, we take $r=4$ instead of $r=3$ for
reconstructing the state via Theorem \ref{Theorem2}. When the singular
matrix error problem occurs, we can solve the problem via increasing the value of $r$ for reconstructing the state.
\label{Remark13}
\end{remark}
\section{Numerical Examples \label{sectionNE}}
In this section, the correctness and effectiveness of the proposed state reconstruction methods are illustrated by an example of four-dimensional dynamic systems and a real-world example of three-inertia systems.
\subsection{ Example of Four-Dimensional Dynamic Systems \label{sectionNEa}}
\par
Consider the four-dimensional model described by (\ref{sxk}) and (\ref{syk}) where $q=6$ and
\begin{align}
&A=\!\left[\!\!
    \begin{array}{cccc}
      2.3 & -0.6 & 3.8 & 0.4 \\
      3.2 & -1.6 & 0.7 & 0.4 \\
      1.7 & 2.8 & 5.2 & 4.3 \\
      -3.1 & 2.4 & 3.7 & 4.8 \\
    \end{array}\!\!
  \right]\!\!,\ B=\!\left[\!\!
                    \begin{array}{c}
                      1 \\
                      1 \\
                      1 \\
                      1 \\
                    \end{array}
                 \!\! \right],\ u_k=3.6,
\nonumber\\
&C=\left[
     \begin{array}{cccc}
       1 & 0 & 0 & 1 \\
       1 & 0 & 1 & 0 \\
       1 & 1 & 0 & 0 \\
       0 & 1 & 1 & 0 \\
       0 & 1 & 0 & 1 \\
       0 & 0 & 1 & 1 \\
     \end{array}
   \right],\ a_k=\!\left[\!\!
                   \begin{array}{c}
                     2000+k/(k+1) \\
                     0 \\
                     3000+k/(k+2) \\
                     1500\textrm{sin}(2k+1) \\
                     0 \\
                     3000\textrm{cos}(2k+3) \\
                   \end{array}
                \!\! \right].
\nonumber
 \end{align}
 For the initial state $x_0$, we take the following four cases with different\vspace{2pt} values:\\
\textbf{Case 1.} $x_0=\!\left[\!\!
                   \begin{array}{c}
                     25.2 \\
                     -16.2 \\
                     123.3 \\
                     4.9 \\
                   \end{array}
                 \!\!\right]$,
\textbf{Case 2.} $x_0=\!\left[\!\!
                   \begin{array}{c}
                     -25.2 \\
                     36.8 \\
                     -123.3 \\
                     -4.9 \\
                   \end{array}
                 \!\!\right]$,\vspace{3pt}\\
\textbf{Case 3.} $x_0=\!\left[\!\!
                   \begin{array}{c}
                     382.4 \\
                     739.2 \\
                     371.8 \\
                     371.2 \\
                   \end{array}
                 \!\!\right]$,
\textbf{Case 4.} $x_0=\!\left[\!\!
                   \begin{array}{c}
                     4.2 \\
                     -2.5 \\
                     -4.2 \\
                     7.2 \\
                   \end{array}
                 \!\!\right]$.\vspace{1pt}
\par
We first test the correctness and effectiveness of the
proposed methods for reconstructing the initial state $x_0$ given in Case 1.
Since the maximum number of attacked sensors is four, we have $s=4$. Then,
we easily see that the system considered in the example is $(s+1)$-sparse
observable with $s=4$ and satisfies the condition of Theorem \ref{Theorem2} when $r\geq4$.
The selection of $\Lambda_j$ with $j=1,2,\cdots,C_q^{s+1}$ is according to common sequential order, that is,
$\Lambda_1=\{1,2,3,4,5\}$, $\Lambda_2=\{1,2,3,4,6\}$, $\Lambda_3=\{1,2,3,5,6\}$, $\Lambda_4=\{1,2,4,5,6\}$, $\Lambda_5=\{1,3,4,5,6\}$ and $\Lambda_6=\{2,3,4,5,6\}$.
Using Theorem \ref{Theorem2} with $r=4$ and the measurements from time steps $0$ to $3$, we obtain
$Y(3,4)=\big\{y_{0}^{(j)}|j=1,2,\cdots,6\big\}$ with
\begin{align}
&y_{0}^{(1)}=\!\!\left[\!\!
                   \begin{array}{c}
                     3403.72 \\
                     6226.33 \\
                     -609.49 \\
                     -2232.29 \\
                   \end{array}
                 \!\!\right],y_{0}^{(2)}=\!\!\left[\!\!
                   \begin{array}{c}
                     25.2 \\
                     -16.2 \\
                     123.3 \\
                     4.9 \\
                   \end{array}
                 \!\!\right],y_{0}^{(3)}=\!\!\left[\!\!
                   \begin{array}{c}
                     431.81 \\
                     1354.19 \\
                     15.12 \\
                     -588.06 \\
                   \end{array}
                 \!\!\right],\nonumber\\
&y_{0}^{(4)}=\!\!\left[\!\!
                   \begin{array}{c}
                     1641.01 \\
                     1367.99 \\
                     -443.54 \\
                     -358.42 \\
                   \end{array}
                 \!\!\right],y_{0}^{(5)}=\!\!\left[\!\!
                   \begin{array}{c}
                     25.2 \\
                     -16.2 \\
                     123.3 \\
                     4.9 \\
                   \end{array}
                 \!\!\right],y_{0}^{(6)}=\!\!\left[\!\!
                   \begin{array}{c}
                     14837.15 \\
                     7180.53 \\
                     9125.71 \\
                     -12807.05 \\
                   \end{array}
                 \!\!\right].\nonumber
 \end{align}
We easily see that there exists $q-s=2$ elements  $y_{0}^{(2)}$ and $y_{0}^{(5)}$ with the same value in the set $Y(3,4)$.
Then, noting that the remaining elements of the set $Y(3,4)$ are not equal to\vspace{2pt} each other, we should obtain
$x_0=y_{0}^{(2)}=y_{0}^{(5)}$ if Theorem \ref{Theorem2} is correct. From the numerical results, we have $x_0=y_{0}^{(2)}=y_{0}^{(5)}$ which means the
correctness of Theorem \ref{Theorem2}. Also, we can obtain the state $x_0$ by searching elements with the same value in the set $Y(3,4)$, which indicates the  effectiveness
of Theorem \ref{Theorem2}. We easily conclude that the system considered in the example is $s$-sparse
observable and the condition for running Algorithm 1 is satisfied when $r\geq2$. The selection of $\Gamma_j$ with $j=1,2,\cdots,C_q^s$\vspace{2pt} is according to common sequential order, that is,
$\Gamma_1=\{1,2,3,4\}$, $\Gamma_2=\{1,2,3,5\}$, $\cdots$, $\Gamma_{15}=\{3,4,5,6\}$.\vspace{2pt} Considering computation error made by MATLAB, we use\vspace{3pt} $\Big\|x_{k-r+2}^{(j)}-Ax_{k-r+1}^{(j)}-Bu_{k-r+1}\Big\|_{2}\leq 0.1$ instead of
$x_{k-r+2}^{(j)}-Ax_{k-r+1}^{(j)}-Bu_{k-r+1}=0$\vspace{2pt} in running  Algorithm 1.
Using Algorithm 1 with $r=2$ and the\vspace{2pt} measurements from time steps $0$ to $2$, we obtain
$\check{X}_1(1,2)=\{x_0^{(8)}\}$\vspace{2pt} where $x_0^{(8)}=\!\!\left[\!\!
                                                    \begin{array}{cccc}
                                                      25.2  & -16.2 & 123.3 & 4.9 \\
                                                    \end{array}
                                                  \!\!\right]^\textrm{T}$.
Noting that $x_0=x_0^{(8)}$ and the set $\check{X}_1(1,2)$\vspace{2pt} has one and only one element $x_0^{(8)}$, we verify the correctness and effectiveness of Algorithm 1.
In the same way, we can reconstruct the state $x_0$ for the remaining cases using both Theorem \ref{Theorem2} and  Algorithm 1, which shows the correctness and effectiveness of the two proposed state reconstruction methods.
Noticing $s>\lceil q/2\rceil$ at $s=4$ and $q=6$,
the numerical results also show that the two proposed state reconstruction methods still can reconstruct the state in $s\geq\lceil q/2\rceil$ while the previous methods cannot reconstruct the state in $s\geq\lceil q/2\rceil$.
\par
To further confirm the correctness and effectiveness of the proposed methods, we change the value of $a_k$ into
\begin{align}
a_k=\!\left[\!\!
                   \begin{array}{c}
                     2500+k/(k+1) \\
                     3000+k/(k+2) \\
                     0 \\
                     4500\textrm{sin}(3k+1) \\
                     3500+k/(k+3) \\
                     -3000\textrm{cos}(k+4) \\
                   \end{array}
                \!\! \right]\nonumber
  \end{align}
and we only select the initial state $x_0$ given in Case 1. We easily obtain $s=5$ and the condition of $(s+1)$-sparse observability is not satisfied.
Hence, we cannot reconstruct the state $x_0$ using the state reconstruction method based on SESVS presented in Section \ref{secmrb}.
Since the condition of $s$-sparse observability is satisfied, we can use Algorithm 1 for reconstructing the state $x_0$.
Using\vspace{2pt} Algorithm  1 with $r=4$ and the measurements from time steps\vspace{3pt} $0$ to $4$,
we get $\check{X}_1(3,4)=\{x_0^{(4)}\}$ with\vspace{2pt} $x_0^{(4)}=\!\!\left[\!\!
                                                    \begin{array}{cccc}
                                                      25.2  & -16.2 & 123.3 & 4.9 \\
                                                    \end{array}
                                                  \!\!\right]^\textrm{T}$.
Noticing $x_0=x_0^{(4)}$,\vspace{2pt} we confirm the correctness and effectiveness of Algorithm 1.
\subsection{ Example of Three-Inertia Systems \label{sectionNEb}}
Consider a three-inertia system \cite{nu1}, \cite{nu2} and its dynamic equation in discrete time can be described by (\ref{sxk}) after the discretization for (20) of \cite{nu1}
using the forward Euler method
where\vspace{0.1pt}
\begin{align}
&A\!=\!\!\left[\!\!\!\!
    \begin{array}{cccccc}
      1 & T & 0 & 0 & 0 & 0\vspace{2pt} \\
      -\frac{k_1 T}{J_1} & 1-\frac{b_1 T}{J_1} & \frac{k_1 T}{J_1} & 0 & 0 & 0\vspace{2pt} \\
      0 & 0 & 1 & T & 0 & 0\vspace{2pt} \\
      \frac{k_1 T}{J_2} & 0 & -\frac{k_1+k_2}{J_2}T  & 1-\frac{b_2 T}{J_2} & \frac{k_2 T}{J_2} & 0\vspace{2pt} \\
      0 & 0 & 0 & 0 & 1 & T\vspace{2pt} \\
      0 & 0 & \frac{k_2 T}{J_3} & 0 & -\frac{k_2 T}{J_3} & 1-\frac{b_3 T}{J_3}\vspace{2pt} \\
    \end{array}
  \!\!\!\!\right]\!\!,
\nonumber\\
& x_k=\!\!\left[\!\!
        \begin{array}{c}
          \theta_{1,k}\vspace{1pt} \\
          \dot{\theta}_{1,k}\vspace{1pt} \\
          \theta_{2,k}\vspace{1pt} \\
          \dot{\theta}_{2,k}\vspace{1pt} \\
          \theta_{3,k}\vspace{1pt} \\
          \dot{\theta}_{3,k}\vspace{1pt} \\
        \end{array}
     \!\! \right], B=\!\!\left[\!\!
        \begin{array}{c}
          0\vspace{1pt} \\
          \frac{ T}{J_1}\vspace{1pt} \\
          0\vspace{1pt} \\
          0\vspace{1pt} \\
          0\vspace{1pt} \\
          0\vspace{1pt} \\
        \end{array}
     \!\! \right], u_k=T_{m,k};
\nonumber
  \end{align}
$k_j$ with $j=1,2$ denotes the torsional stiffness of the $j$th shaft; $T$ is the sampling
period in seconds; $J_1$ is the inertia of the drive motor; $J_2$ is the inertia of the middle body;
$J_3$ is the inertia of the load; $b_j$ with $j=1,2,3$ is the mechanical damping of the $j$th inertia;
$\theta_j$ with\vspace{1pt} $j=1,2,3$ is the
absolute angular position of the $j$th inertia; $\dot{\theta}_{j}$ with $j=1,2,3$ is the speed of the $j$th inertia;
and $T_m$ is the motor drive torque.
The output measurements can be expressed as (\ref{syk})
where\vspace{-0.1pt}
\begin{align}
&C=\left[
     \begin{array}{cccccc}
       1 & 0 & 0 & 0& 0 & 0 \\
       0 & 0 & 1 & 0& 0 & 0 \\
       0 & 0 & 0 & 0& 1 & 0 \\
       1 & 0 & -1 & 0& 0 & 0 \\
       1 & 0 & 0 & 0& -1 & 0 \\
       1 & 0 & 1 & 0& -1 & 0 \\
       1 & 0 & -1 & 0& 1 & 0 \\
     \end{array}
   \right].
\nonumber
 \end{align}
Here, $k_1=k_2=1.38 $ N$\cdot$m/rad,\vspace{1pt} $T=0.005$ s, $J_1=0.01$ kg$\cdot$m$^2$, $J_2=0.02$ kg$\cdot$m$^2$,\vspace{1pt}
$J_3=0.03$ kg$\cdot$m$^2$, $b_1=b_2=b_3=0.006$ N$\cdot$m/(rad/s) and $T_{m,k}=9.5+0.1\textrm{sin}k$ N$\cdot$m.\vspace{1pt}
The initial state is chosen as $x_0=\!\!\left[\!\!
                                                    \begin{array}{cccccc}
                                                      0.2  & 1.2 & 0.19 & 1.1 & 0.3 & 1.6 \\
                                                    \end{array}
                                                  \!\!\right]^\textrm{T}$. The first four sensors are attacked with\vspace{1pt}
$a_{k,1}=125+\sqrt{k}$, $a_{k,2}=143\textrm{cos}k$, $a_{k,3}=234\textrm{sin}(k+1)-143\textrm{cos}k$ and $a_{k,4}=16-80\textrm{sin}(k+3)$.
\par
The correctness and effectiveness of the theoretical results are examined by reconstructing the initial state $x_0$.
Noticing $s=4$ and $q=7$, we easily see that the condition required by Theorem \ref{Theorem2} is satisfied when $r\geq3$.
In order to avoid singular matrix error in running MATLAB,
we select $r=4$ instead of $r=3$.
Using Theorem \ref{Theorem2}\vspace{2pt} and the measurements from time steps 0 to 3, we get $Y(3,4)=\big\{y_{0}^{(j)}|j=1,2,\cdots,21\big\}$ with\vspace{-0.1pt}
\begin{align}
&y_{0}^{(1)}=\!\!\left[\!\!
              \begin{array}{c}
                0.2 \\
                1.2 \\
                0.19 \\
                1.1 \\
                0.3 \\
                1.6 \\
              \end{array}
            \!\!\right],
            y_{0}^{(2)}=\!\!\left[\!\!
              \begin{array}{c}
                0.2 \\
                1.2 \\
                0.19 \\
                1.1 \\
                0.3 \\
                1.6 \\
              \end{array}
            \!\!\right],
            y_{0}^{(3)}=\!\!\left[\!\!
              \begin{array}{c}
                0.2 \\
                1.2 \\
                0.19 \\
                1.1 \\
                0.3 \\
                1.6 \\
              \end{array}
            \!\!\right],\nonumber\\
&y_{0}^{(4)}=\!\!\left[\!\!
              \begin{array}{c}
                -5962.39 \\
                6415563.29 \\
                -5979.7 \\
                6404824.55 \\
                -31.27 \\
                -3187.18 \\
              \end{array}
            \!\!\right],\cdots,
            y_{0}^{(21)}=\!\!\left[\!\!
              \begin{array}{c}
                125.38 \\
                115.18 \\
                143.19 \\
                -13146.2 \\
                -41075.95 \\
                14540631.71 \\
              \end{array}
            \!\!\right]
\nonumber
 \end{align}
where the remaining elements of the set\vspace{2pt} $Y(3,4)$ are not equal to each other except for $y_{0}^{(1)}$, $y_{0}^{(2)}$
and $y_{0}^{(3)}$. From the above numerical results, we see that there\vspace{2pt} exists $q-s$ elements $y_{0}^{(1)}$, $y_{0}^{(2)}$
and $y_{0}^{(3)}$ with $q=7$ and $s=4$\vspace{2pt} such that $x_0=y_{0}^{(1)}=y_{0}^{(2)}=y_{0}^{(3)}$, which means the correctness of Theorem \ref{Theorem2}.
Since the remaining elements of the set\vspace{2pt} $Y(3,4)$ are not equal to each other, we can find the state $x_0$ by searching elements with the same value in the set
$Y(3,4)$. Hence, Theorem \ref{Theorem2} is effective for reconstructing the state $x_0$.
We easily see that the condition required by Algorithm 1 is satisfied when $r\geq 2$. Using Algorithm 1 with $r=2$ and the measurements from
time steps 0 to 2, we derive $\check{X}_1(1,2)=\{x_0^{(1)}\}$\vspace{2pt} with $x_0^{(1)}=\!\!\left[\!\!
                                                    \begin{array}{cccccc}
                                                      0.2  & 1.2 & 0.19 & 1.1 & 0.3 & 1.6 \\
                                                    \end{array}
                                                  \!\!\right]^\textrm{T}.$ Since $x_0=x_0^{(1)}$ and there is only one element $x_0^{(1)}$ in the set $\check{X}_1(1,2)$, the correctness and effectiveness of Algorithm 1
are verified.
\section{Conclusion \label{sectionCo}}
When some of the
sensors can be arbitrarily corrupted, it has been proved that
the state is $s$-error correctable if the system under consideration is $s$-sparse
observable. Two secure state reconstruction methods have been proposed where the first method is based on SESVS and the second method is based on SESGC.
As long as proper conditions are satisfied, the two proposed methods
 can accurately reconstruct the state even
though half or more of all measurement
sensors are arbitrarily corrupted
by malicious attacks. Moreover, after establishing and analyzing the conditions that the two proposed methods fail to reconstruct the state,  it has been concluded that
 it is very hard to prevent the state reconstruction when either of the two proposed methods is used. The numerical results for
an example of four-dimensional dynamic systems and a real-world example of three-inertia systems have shown
 the correctness and effectiveness of the two proposed methods.
% if have a single appendix:
%\appendix[Proof of the Zonklar Equations]
% or
%\appendix  % for no appendix heading
% do not use \section anymore after \appendix, only \section*
% is possibly needed

% use appendices with more than one appendix
% then use \section to start each appendix
% you must declare a \section before using any
% \subsection or using \label (\appendices by itself
% starts a section numbered zero.)
 \appendices
\section{Proof of Lemma \ref{Lemma1} and Theorem \ref{Theorem1} \label{appenT1}}
\subsection {Proof of Lemma \ref{Lemma1} \label{appensL1}}
From (\ref{ykGd}), it follows that\vspace{-0.1pt}
\begin{align}
O_{\Gamma}x_{k-r+1}=Y_{k,\Gamma}-D_{\Gamma}U_{k-1}.
\nonumber
\end{align}
 Multiplying both sides of the above equality by $O_{\Gamma}^\textrm{T}$, we obtain
\begin{align}
O_{\Gamma}^\textrm{T}O_{\Gamma}x_{k-r+1}=O_{\Gamma}^\textrm{T}(Y_{k,\Gamma}-D_{\Gamma}U_{k-1}).
\label{PLemm1a}
\end{align}
 Using the given condition that $\big(A,C(\Gamma)\big)$ is observable, we see that
there exist a $r$ with $r\leq n$ such that the rank of $O_{\Gamma}$ is $n$, that is, rank$(O_{\Gamma})=n$ where
$O_{\Gamma}$ is defined in (\ref{ykGe}). Hence, the minimum value of $r$ such that the rank of $O_{\Gamma}$ is $n$ exists and is less than or equal to $n$, which means $a\leq n$.
Since rank$(M^\textrm{T}M)=$rank$(M)$ for any matrix $M$, we conclude that
the rank of $O_{\Gamma}^\textrm{T}O_{\Gamma}$ is $n$ at $r=a$. From the definition of\vspace{2pt} $O_{\Gamma}$, we easily see that
 $O_{\Gamma}^\textrm{T}O_{\Gamma}$ is a $n\times n$ matrix. Then, using the result rank$(O_{\Gamma}^\textrm{T}O_{\Gamma})=n$, we
see that $O_{\Gamma}^\textrm{T}O_{\Gamma}$ is a full-rank square matrix. Hence, $O_{\Gamma}^\textrm{T}O_{\Gamma}$ is invertible.
Then,\vspace{2pt} multiplying both sides of (\ref{PLemm1a}) by $(O_{\Gamma}^\textrm{T}O_{\Gamma})^{-1}$, we get
\begin{align}
x_{k-r+1}=(O_{\Gamma}^\textrm{T}O_{\Gamma})^{-1}O_{\Gamma}^\textrm{T}(Y_{k,\Gamma}-D_{\Gamma}U_{k-1}).
\nonumber
\end{align}
Then, replacing $(O_{\Gamma}^\textrm{T}O_{\Gamma})^{-1}O_{\Gamma}^\textrm{T}$ by $L_\Gamma$, we obtain
(\ref{Lem1a}) at $r=a$, which means that the lemma holds at $r=a$.
We easily see that $O_{\Gamma}^\textrm{T}O_{\Gamma}$ is invertible in $r>a$. Then, making reference to the corresponding proof of Lemma \ref{Lemma1} at $r=a$, we can easily prove
the lemma holds at $r>a$. Then, noting that the lemma holds at $r= a$, we see that the lemma holds in $r\geq a$.
This completes the proof of the lemma.
\subsection {Proof of Theorem \ref{Theorem1} \label{appensT1}}
In order to prove Theorem \ref{Theorem1}, we first derive the following two lemmas.
\begin{lemma} Let $\Gamma(q,s)$ denote the set of all possible cases of index set $\Gamma$ defined in Remark \ref{Remark2}\vspace{1pt} with $\Gamma\subseteq\{1,2,\cdots,q\}$ and $|\Gamma|= s$,
that is, $\Gamma(q,s)=\{\Gamma_j|j=1,2,\cdots,C_q^s\}$.
Then,
if the system under consideration is $s$-sparse
observable,
for $\forall\Gamma\in\Gamma(q,s)$, there exists\vspace{1pt} a positive integer $b$ with $b=\underset{_{j\in \{1,2,\cdots,C_q^s\}}}{\textrm{max}}r_j\leq n$ such that
the state $x_{k-r+1}$ can be exactly computed via
\begin{align}
x_{k-r+1}=L_\Gamma(Y_{k,\Gamma}-D_{\Gamma}U_{k-1})
\nonumber
\end{align}
at $r= b$ where $r_j$ is the minimum value of $r$ such that the rank of $O_{\Gamma_j}$ is $n$.
 \label{LeThe1}
\end{lemma}
\textit{Proof:} In order to prove the lemma, we first need to prove $b\leq n$. Then, noting that there are $C_q^s$ possible cases for $\Gamma$, we need to prove that the statement holds for all $C_q^s$ cases,
that is, the statement holds for $\Gamma=\Gamma_i$ with $i=1,2,\cdots,C_q^s$.
We first prove $b\leq n$.
Making reference to the proof of Lemma \ref{Lemma1}, we get $r_i\leq n$ with $i=1,2,\cdots,C_q^s$ if $\big(A,C(\Gamma_i)\big)$ is observable.
We easily see from Definition \ref{definition1} that $\big(A,C(\Gamma_i)\big)$ is observable if the system under consideration is s-sparse
observable. Hence, we conclude that $r_i\leq n$ with $i=1,2,\cdots,C_q^s$ if the system under consideration is s-sparse
observable.
 Then, noticing $b=\underset{_{j\in \{1,2,\cdots,C_q^s\}}}{\textrm{max}}r_j$, we get $ b\leq n$ if the system under consideration is s-sparse
observable.
Then, we prove that the statement holds for all $C_q^s$ cases.
For the $i$th possible case, that is, $\Gamma=\Gamma_i$, we see from Lemma \ref{Lemma1} and the proof of Lemma \ref{Lemma1} that $x_{k-r+1}=L_{\Gamma_i}(Y_{k,\Gamma_i}-D_{\Gamma_i}U_{k-1})$
in $r\geq r_i$\vspace{1pt} with $r_i\leq n$ if
$\big(A,C(\Gamma_i)\big)$ is observable.\vspace{2pt}  Then, noting that  $\big(A,C(\Gamma_i)\big)$ is observable if the system under consideration is s-sparse
observable, we have  $x_{k-r+1}=L_{\Gamma_i}(Y_{k,\Gamma_i}-D_{\Gamma_i}U_{k-1})$
in $r\geq r_i$ if the system under consideration is s-sparse
observable.
Then, noticing $b\geq r_i$, we get $x_{k-r+1}=L_{\Gamma_i}(Y_{k,\Gamma_i}-D_{\Gamma_i}U_{k-1})$
at $r= b$ if the system under consideration is s-sparse
observable. Since the above statement holds for any $i$, noticing
$b\leq n$, we prove that the statement holds for all $C_q^s$ cases, which means that the lemma holds.\vspace{1pt}
\begin{lemma}
If the system under consideration is $s$-sparse
observable, any element in the set $X(k,r)$ exists at $r= b$.
 \label{LeThe1a}
\end{lemma}
\textit{Proof:} Obviously,\vspace{1pt} we need to prove $x_{k-r+1}^{(j)}$ with $j=1,2,\cdots,C_q^s$ exists at $r= b$.
If $\big(A,\Gamma_j\big)$ is observable, making reference to the proof of Lemma \ref{Lemma1}, we see that $O_{\Gamma_j}^\textrm{T}O_{\Gamma_j}$ is invertible at\vspace{1pt} $r= b$, which means that $x_{k-r+1}^{(j)}$ exists at $r= b$ by
using (\ref{Lem1ana}) and noticing $x_{k-r+1}^{(j)}=L_{\Gamma_j}(Y_{k,\Gamma_j}-D_{\Gamma_j}U_{k-1})$.
Then,
noting that $\big(A,\Gamma_j\big)$ is observable if the system under consideration is $s$-sparse
observable,
we see that $x_{k-r+1}^{(j)}$ exists at $r= b$ if the system under consideration is $s$-sparse
observable. Since the above statement holds for any $j$, we prove the lemma.
\par
We now provide a proof of Theorem \ref{Theorem1}.
For the $i$th cases of $\Gamma$, namely, $\Gamma=\Gamma_i$,
using Lemma \ref{LeThe1}, we get
 $x_{k-r+1}=L_{\Gamma_i}(Y_{k,\Gamma_i}-D_{\Gamma_i}U_{k-1})$ at $r=b$ if the system under consideration is s-sparse
observable where $b$ with $b\leq n$ is defined in Lemma \ref{LeThe1}.\vspace{2pt} Then, at $r=b$ and $\Gamma=\Gamma_i$, we get $x_{k-r+1}=x_{k-r+1}^{(i)}$ by\vspace{2pt} noticing $x_{k-r+1}=L_{\Gamma_i}(Y_{k,\Gamma_i}-D_{\Gamma_i}U_{k-1})$
and $x_{k-r+1}^{(i)}=L_{\Gamma_i}(Y_{k,\Gamma_i}-D_{\Gamma_i}U_{k-1})$\vspace{1pt} if the system under consideration is s-sparse
observable. \vspace{2pt} Then, using Lemma \ref{LeThe1a} and noticing $x_{k-r+1}^{(i)}\in X(k,r)$,\vspace{1pt} we conclude that $x_{k-r+1}$ is equal to at least an element in the set $X(k,r)$ at $r=b$ and $\Gamma=\Gamma_i$ if the system under consideration is s-sparse
observable.
Since the above statement holds for any $i$, that is, $x_{k-r+1}$ with $r=b$ is equal to\vspace{1pt} at least an element in the set $X(k,r)$ at $\Gamma=\Gamma_i$  with $i=1,2,\cdots,C_q^s$, we conclude that
$x_{k-r+1}$ with $r=b$ is equal to at least an element in the set $X(k,r)$ for all possible cases of $\Gamma$ if the system under consideration is s-sparse
observable, which means that the theorem holds.
\section{Proof of  Theorem  \ref{Theorem2} and Corollary \ref{corollary1}\label{appenT2}}
\subsection {Proof of  Theorem  \ref{Theorem2} \label{appenT21}}
Let $b_1$ denote the $(s+1)$-sparse observable lower bound, that is, $b_1=\underset{_{j\in \big\{1,2,\cdots,C_q^{s+1}\big\}}}{\textrm{max}}r_{1j}$ where $r_{1j}$ is the minimum value of $r$ such that the rank of $O_{\Lambda_j}$ is $n$.
Making reference to the proof of Lemma \ref{LeThe1}, we get $b_1\leq n$ if the system under consideration is $(s+1)$-sparse
observable.
Let $\check{\Gamma}_i=Q\backslash \Gamma_i$ where $Q=\{1,2,\cdots,q\}$ and $\Gamma_i$ is the
 $i$th possible case of $\Gamma$ with $i=1,2,\cdots,C_q^s$.
Making reference to the proof of Lemma \ref{LeThe1a}, we see that any element in the set $Y(k,r)$ exists at $r= b_1$ if the system\vspace{1pt} under consideration is $(s+1)$-sparse
observable.
 Noticing $|Q|=q$ and $|\Gamma_i|=s$,\vspace{2pt} we have $|\check{\Gamma}_i|=q-s$. From the given condition, we have
$|\check{\Gamma}_i|\geq 2$.\vspace{2pt} Let $\check{\Gamma}_{il}$ denote the $l$th element of $\check{\Gamma}_i$\vspace{2pt} with $l=1,2,\cdots,q-s$ and let
$\Theta_{il}=\Gamma_i\cup\{\check{\Gamma}_{il}\}$. Then, we easily conclude that $Q\backslash\Theta_{il}$ does not contain any attacked sensor, that is, any element in $Q\backslash\Theta_{il}$ does not belong to the set of attacked sensors.
Hence, noting that the sensor cannot be attacked when it does not belong to the set of attacked sensors, we see from
(\ref{ykGa})$-$(\ref{ykGd})
that
\begin{align}
Y_{k,\Theta_{il}}=O_{\Theta_{il}}x_{k-r+1}+D_{\Theta_{il}}U_{k-1}.
\nonumber
\end{align}
Then, making reference to the proof of Lemma \ref{Lemma1},  we see that if $\big(A,C(\Theta_{il})\big)$ is observable, there exists a $\breve{r}_{il}\leq n$ satisfying
\begin{align}
x_{k-r+1}=x_{k-r+1}^{(il)}\nonumber
\end{align}
in $r\geq \breve{r}_{il}$
where
\begin{align}
x_{k-r+1}^{(il)}=L_{\Theta_{il}}(Y_{k,\Theta_{il}}-D_{\Theta_{il}}U_{k-1})\label{Ptheo2c}
\end{align}
and $\breve{r}_{il}$ is the minimum value of $r$ such that the rank of $O_{\Theta_{il}}$ is $n$.
From the definition $\Theta_{il}=\Gamma_i\cup\{\check{\Gamma}_{il}\}$, we have $\Theta_{il}\subseteq Q$ with $|\Theta_{il}|=s+1$, which mean
\vspace{2pt} $\Theta_{il}\in \{\Lambda_j|j=1,2,\cdots,C_q^{s+1}\}$. Hence, we get $\breve{r}_{il}\leq b_1$.
Then, noticing $x_{k-r+1}=x_{k-r+1}^{(il)}$ in $r\geq \breve{r}_{il}$ with $\breve{r}_{il}\leq b_1$ if $\big(A,C(\Theta_{il})\big)$ is observable, we see that $x_{k-r+1}$ is equal to any element in the
set
$X^{(i)}(k,r)=\big\{x_{k-r+1}^{(il)}|l=1,2,\cdots,q-s\big\}$ at $r=b_1$ if $\big(A,C(\Theta_{il})\big)$ is observable for any $l$ where
$x_{k-r+1}^{(il)}$ is obtained from (\ref{Ptheo2c}).\vspace{2pt}
Noticing $\Theta_{il}\subseteq Q$ with $|\Theta_{il}|=s+1$,\vspace{2pt} we conclude that
$C(\Theta_{il})$ is obtained from deleting $s+1$ rows\vspace{2pt} of $C$. Hence, $\big(A,C(\Theta_{il})\big)$ is observable for any $l$ if the system under consideration is $(s+1)$-sparse
observable satisfying $q\geq s+2$. Hence, if the system  under consideration is $(s+1)$-sparse
observable satisfying $q\geq s+2$, $x_{k-r+1}$ is equal\vspace{2pt} to all the $q-s$ elements in the
set
$X^{(i)}(k,r)$ at $r=b_1$. Noticing $\Theta_{il}\subseteq Q$ with $|\Theta_{il}|=s+1$, we obtain\vspace{2pt} $\{\Theta_{il}|l=1,2,\cdots,q-s\}\subseteq \{\Lambda_j|j=1,2,\cdots,C_q^{s+1}\}$, which means $X^{(i)}(k,r)\subseteq Y(k,r)$
where any element in the set $Y(k,r)$ exists at $r= b_1$ under the condition of  $(s+1)$-sparse observability. Hence,
if the system under consideration is $(s+1)$-sparse
observable satisfying $q\geq s+2$, $x_{k-r+1}$ with $r=b_1$ is equal to at least $q-s$ elements in the
set
$Y(k,r)$ at $\Gamma=\Gamma_i$. Since the statement holds for any $i$, we prove the theorem.
\subsection {Proof of  Corollary \ref{corollary1} \label{appenT22}}
From the given condition, we get $|\check{\Gamma}_{i}|\geq\tau+1$\vspace{2pt} where $\check{\Gamma}_{i}$ is defined in the proof of Theorem \ref{Theorem2}.
From Remark \ref{Remark2},\vspace{2pt} we see that there are $C_{q-s}^\tau$ possible cases for $\check{\Gamma}_{i}^{[\tau]}$ \vspace{2pt}with $\check{\Gamma}_{i}^{[\tau]}\subseteq \check{\Gamma}_{i}$ and
$\big|\check{\Gamma}_{i}^{[\tau]}\big|=\tau$.
Let $\check{\Gamma}_{il}^{[\tau]}$\vspace{2pt} be the $l$th possible case of $\check{\Gamma}_{i}^{[\tau]}$ with $l=1,2,\cdots,C_{q-s}^\tau$, and let
$\Theta_{il}^{[\tau]}=\Gamma_i\cup\check{\Gamma}_{il}^{[\tau]}$. Then, making reference to the corresponding proof of Theorem  \ref{Theorem2}, we easily prove the corollary.
\section{Proof of  Theorem \ref{Theorem3}  \label{appenT3}}
\textit{Sufficiency:}
Let
\begin{align}
z_{k-r+1}^{(\rho)}=L_{\Upsilon_\rho}(Y_{k,\Upsilon_\rho}-D_{\Upsilon_\rho}U_{k-1}).\label{Pthe3a1}
\end{align}
From the definition of $Y(k,r)$ given in Theorem \ref{Theorem2},\vspace{2pt} we easily obtain $z_{k-r+1}^{(\rho)}\in Y(k,r)$ for any $\rho$.
Hence,\vspace{2pt} we need to prove $z_{k-r+1}^{(1)}=z_{k-r+1}^{(2)}=\cdots=z_{k-r+1}^{(q-s)}\neq x_{k-r+1}$. \vspace{2pt}
Making reference to (\ref{ykGa})$-$(\ref{ykGe}), we have\vspace{-0.1pt}
\begin{align}
Y_{k,\Upsilon_\rho}=O_{\Upsilon_\rho}x_{k-r+1}+D_{\Upsilon_\rho}U_{k-1}+A_{k,\Upsilon_\rho}\label{REm3b}
\end{align}
where $A_{\Upsilon_\rho}$ is defined in (\ref{the3a3}).
Substituting (\ref{REm3b}) into (\ref{Pthe3a1}), we get
\begin{align}
z_{k-r+1}^{(\rho)}= &L_{\Upsilon_\rho}(O_{\Upsilon_\rho}x_{k-r+1}+A_{k,\Upsilon_\rho})\nonumber\\
=&L_{\Upsilon_\rho}O_{\Upsilon_\rho}x_{k-r+1}+L_{\Upsilon_\rho}A_{k,\Upsilon_\rho}\nonumber\\
=&x_{k-r+1}+L_{\Upsilon_\rho}A_{k,\Upsilon_\rho}\label{REm3e}
\end{align}
where the last equality follows from (\ref{Lem1ana}).\vspace{2pt}
Using the condition given in (\ref{the3a2}) and using (\ref{REm3e}), we get $z_{k-r+1}^{(1)}= z_{k-r+1}^{(\rho)}$\vspace{2pt} for any $\rho$.
Utilizing the condition given in (\ref{the3a1}) and using (\ref{REm3e}), we obtain $x_{k-r+1}\neq z_{k-r+1}^{(1)}$.
Hence, using the conditions given in (\ref{the3a2}) and (\ref{the3a1}), and using (\ref{REm3e}), we get $z_{k-r+1}^{(1)}= z_{k-r+1}^{(\rho)}$ for any $\rho$ and $x_{k-r+1}\neq z_{k-r+1}^{(1)}$.
Then,\vspace{2pt} noting that $z_{k-r+1}^{(1)}=z_{k-r+1}^{(2)}=\cdots=z_{k-r+1}^{(q-s)}\neq x_{k-r+1}$ if $z_{k-r+1}^{(1)}= z_{k-r+1}^{(\rho)}$\vspace{2pt} for any $\rho$ and $x_{k-r+1}\neq z_{k-r+1}^{(1)}$,
we prove the sufficiency.\vspace{2pt}
\\
\textit{Necessity:}
From (\ref{REm3e}), we see that $L_{\Upsilon_1}A_{k,\Upsilon_1}
=L_{\Upsilon_\rho}A_{k,\Upsilon_\rho}$\vspace{2pt} for any $\rho$ and $L_{\Upsilon_1}A_{k,\Upsilon_1}\neq0$ if
$z_{k-r+1}^{(1)}=z_{k-r+1}^{(2)}=\cdots=z_{k-r+1}^{(q-s)}\neq x_{k-r+1}$.\vspace{2pt}
This means that the conditions given in (\ref{the3a2}) and (\ref{the3a1}) hold if there are at least $q-s$ elements in
the set $Y(k,r)$ such that the elements have the same value that is not equal to the state $x_{k-r+1}$.
 \vspace{2pt}
The proof of the necessity is completed.
\section{Proof of Algorithm 1 and Theorem \ref{Theorem4} \label{appenA1}}
\subsection{Proof of Algorithm 1 \label{appenA1a}}
In order to obtain Algorithm 1, we first prove the following two lemmas.
\begin{lemma}
If there\vspace{2pt} exists at least a $\mu$ such that
 $x_{k-r+1}^{(\mu)}=x_{k-r+1}$, it holds that $x_{k-r+\varsigma+2}^{(\mu)}-Ax_{k-r+\varsigma+1}^{(\mu)}\vspace{2pt}-Bu_{k-r+\varsigma+1}=0$, $\varsigma=1,2,\cdots$.\vspace{2pt}
 \label{LeA1b1}
\end{lemma}
\textit{Proof:}\vspace{2pt} Using the given conditions $x_{k-r+1}^{(\mu)}=x_{k-r+1}$, we obtain $x_{k-r+\varsigma+1}^{(\mu)}=x_{k-r+\varsigma+1}$ and
 $x_{k-r+\varsigma+2}^{(\mu)}=x_{k-r+\varsigma+2}$.\vspace{2pt}
  Then, making reference to the proof of Lemma \ref{LeSPEa}, we prove the statement.\vspace{2pt}
\begin{lemma} If the system under consideration is $s$-sparse
observable,
there exists at least a $\mu$ such that
$\mu\in D_\varsigma$, $\varsigma=1,2,\cdots$, and $x_{k-r+1}^{(\mu)}=x_{k-r+1}$.\vspace{2pt}
 \label{LeA1b}
\end{lemma}
\textit{Proof:} We prove the lemma using mathematical induction.\vspace{2pt}  We first prove Lemma \ref{LeA1b} holds at $\varsigma=1$.
\vspace{2pt} From Step 1 of Algorithm 1, we see that $\check{X}_1(k,r)$\vspace{2pt} is obtained from $X(k,r)=\big\{x_{k-r+1}^{(j)}|j=1,2,\cdots,C_q^s\big\}$\vspace{2pt} by collecting $\forall x_{k-r+1}^{(j)}$ such that $ j\in \{1,2,\cdots,C_q^s\}$
and
$x_{k-r+2}^{(j)}-Ax_{k-r+1}^{(j)}-Bu_{k-r+1}=0$.\vspace{2pt} Then,
noting that there exists at least a $\mu$ such that $\mu \in \{1,2,\cdots,C_q^s\}$, $x_{k-r+1}^{(\mu)}=x_{k-r+1}$ and $x_{k-r+2}^{(\mu)}-Ax_{k-r+1}^{(\mu)}\vspace{2pt}-Bu_{k-r+1}=0$ obtained from Lemma \ref{LeSPEa}, we see that
there\vspace{2pt} exists at least a $\mu$ such that $\mu\in D_1$,
$x_{k-r+1}^{(\mu)}=x_{k-r+1}$ and $x_{k-r+2}^{(\mu)}-Ax_{k-r+1}^{(\mu)}-Bu_{k-r+1}=0$.\vspace{2pt} Hence, Lemma \ref{LeA1b}
holds at $\varsigma=1$.\\
Next, we prove Lemma \ref{LeA1b} holds at $\varsigma=\rho+1$ under the assumption that Lemma \ref{LeA1b} holds at $\varsigma=\rho$.
\vspace{2pt}
From Step 4 of Algorithm 1,\vspace{2pt} it follows that $\check{X}_{\rho+1}(k,r)$ is obtained from $\check{X}_{\rho}(k,r)=\big\{x_{k-r+1}^{(j)}|j\in D_\rho\big\}$ by collecting $\forall x_{k-r+1}^{(j)}$
\vspace{2pt} satisfying $j\in D_{\rho}$ and $x_{k-r+\rho+2}^{(j)}-Ax_{k-r+\rho+1}^{(j)}-Bu_{k-r+\rho+1}=0$.\vspace{2pt} Then,
noting that there exists at least a $\mu$\vspace{2pt} such that $\mu\in D_\rho$, $x_{k-r+1}^{(\mu)}=x_{k-r+1}$ and $x_{k-r+\rho+2}^{(\mu)}-Ax_{k-r+\rho+1}^{(\mu)}\vspace{2pt}-Bu_{k-r+\rho+1}=0$ obtained from Lemma \ref{LeA1b1} and the assumption that Lemma \ref{LeA1b} holds at $\varsigma=\rho$,
we prove Lemma \ref{LeA1b} holds at $\varsigma=\rho+1$.
\par
We now provide a proof of Algorithm 1.
From the structure of
Algorithm 1, we see that the purpose of Algorithm 1 is to
search the set $\check{X}_\vartheta(k,r)$\vspace{2pt} with minimal $\vartheta$, $\vartheta=1,2,\cdots,$ such that
there only exists an element in $\check{X}_\vartheta(k,r)$ or each element in $\check{X}_\vartheta(k,r)$ has the same value.
Then, using Algorithm 1, we obtain the state $x_{k-r+1}$ which is equal to any element in the set
$\check{X}_\vartheta(k,r)$. Hence, in order to prove Algorithm 1, \vspace{2pt} we only need to prove that the set $\check{X}_\vartheta(k,r)$ exists.
\vspace{2pt}More precisely, we only need to prove that
 there exists at least a $x_{k-r+1}^{(\mu)}$ with $x_{k-r+1}^{(\mu)}\in \check{X}_\vartheta(k,r)$ such that\vspace{2pt}
$x_{k-r+1}^{(\mu)}=x_{k-r+1}$. Then, using Lemma \ref{LeA1b} and noticing $\check{X}_\vartheta(k,r)=\big\{x_{k-r+1}^{(j)}|j\in D_\vartheta\big\}$, we\vspace{2pt} see that there exists at least a $x_{k-r+1}^{(\mu)}$ with $x_{k-r+1}^{(\mu)}\in \check{X}_\vartheta(k,r)$ such that\vspace{2pt}
$x_{k-r+1}^{(\mu)}=x_{k-r+1}$.
This completes the proof of Algorithm 1.
\subsection{Proof of Theorem \ref{Theorem4} \label{appenA1b}}
The following three lemmas are needed to obtain Theorem \ref{Theorem4}.
\begin{lemma}  When the system under consideration is $s$-sparse
observable, it holds that\vspace{2pt}
$D_{\vartheta}\subseteq D_{\vartheta-1}\subseteq\cdots D_1\subseteq\{1,2,\cdots,C_q^s\}$\vspace{2pt} and $\check{X}_\vartheta(k,r)\subseteq\check{X}_{\vartheta-1}(k,r)\subseteq\cdots\check{X}_1(k,r)\subseteq X(k,r)$.\vspace{2pt}
 \label{LeA1a}
\end{lemma}
\textit{Proof:} From Step 1 of Algorithm 1 and noticing $X(k,r)=\big\{x_{k-r+1}^{(j)}|j=1,2,\cdots,C_q^s\big\}$,\vspace{2pt} we directly obtain
\begin{align}
D_1\subseteq\{1,2,\cdots,C_q^s\},\ \check{X}_1(k,r)\subseteq  X(k,r)\label{PLeA1a1}
\end{align}
 where\vspace{2pt} $x_{k-r+\varsigma}^{(j)}$ exists due to the given condition of $s$-sparse observability.
From Steps 3 and 4 of Algorithm 1,
we get
\begin{align}
D_\varsigma\subseteq D_{\varsigma-1},\ \check{X}_\varsigma(k,r)\subseteq  \check{X}_{\varsigma-1}(k,r)\label{PLeA1a2}
\end{align}
with $\varsigma=2,3,\cdots,\vartheta$. Putting (\ref{PLeA1a1}) and (\ref{PLeA1a2}) together, we prove the lemma.
\begin{lemma} When the system under consideration is $s$-sparse
observable,
for $\forall j\in D_{\varsigma}$,\vspace{2pt} it holds that $x_{k-r+\rho+1}^{(j)}-Ax_{k-r+\rho}^{(j)}-Bu_{k-r+\rho}=0$ with $\rho=1,2,\cdots,\varsigma$.
\vspace{2pt}
 \label{LSPEa1}
\end{lemma}
\par
\textit{Proof:} From the structure of Algorithm 1,\vspace{2pt} we see that
$D_{\varsigma}$ is obtained from searching $\forall j\in D_{\varsigma-1}$
such that\vspace{2pt} $x_{k-r+\varsigma+1}^{(j)}-Ax_{k-r+\varsigma}^{(j)}-Bu_{k-r+\varsigma}=0$ with $D_{0}= \{1,2,\cdots,C_q^s\}$.\vspace{2pt}
Hence, we get $x_{k-r+2}^{(j)}-Ax_{k-r+1}^{(j)}-Bu_{k-r+1}=0$ for $\forall j\in D_{1}$,\vspace{2pt} which means that
Lemma \ref{LSPEa1} holds at $\varsigma=1$.
Similarly,\vspace{2pt} at $\varsigma=2$, we have $x_{k-r+3}^{(j)}-Ax_{k-r+2}^{(j)}-Bu_{k-r+2}=0$\vspace{2pt} for $\forall j\in D_{2}$.
Using the result $D_2\subseteq D_{1}$ obtained\vspace{2pt} from Lemma \ref{LeA1a} and utilizing the above obtained result
$x_{k-r+2}^{(j)}-Ax_{k-r+1}^{(j)}-Bu_{k-r+1}=0$ for $\forall j\in D_{1}$,\vspace{2pt} we have $x_{k-r+2}^{(j)}-Ax_{k-r+1}^{(j)}-Bu_{k-r+1}=0$ for $\forall j\in D_{2}$.
\vspace{2pt}Hence, we get $x_{k-r+2}^{(j)}-Ax_{k-r+1}^{(j)}-Bu_{k-r+1}=0$ and \vspace{2pt} $x_{k-r+3}^{(j)}-Ax_{k-r+2}^{(j)}-Bu_{k-r+2}=0$ for $\forall j\in D_{2}$, which means
that
Lemma \ref{LSPEa1} holds at $\varsigma=2$. Step by step, we prove the lemma.
\begin{lemma}
When the system under consideration is $s$-sparse
observable,
for $\forall j\in D_{\varsigma}$,\vspace{2pt} it holds that  $L_{\Gamma_j}A_{k+\rho,\Gamma_j}-AL_{\Gamma_v}A_{k+\rho-1,\Gamma_j}=0$ with $\rho=1,2,\cdots,\varsigma$.
\vspace{2pt}
 \label{LSPEa2}
\end{lemma}
\textit{Proof:} Making reference to (\ref{ykGa})$-$(\ref{ykGe}), we obtain
\begin{align}
Y_{k,\Gamma_j}=&O_{\Gamma_j}x_{k-r+1}+D_{\Gamma_j}U_{k-1}+A_{k,\Gamma_j}.\label{PThm4b}
\end{align}
Substituting (\ref{PThm4b}) into $x_{k-r+1}^{(j)}=L_{\Gamma_j}(Y_{k,\Gamma_j}-D_{\Gamma_j}U_{k-1})$\vspace{2pt} defined in Theorem \ref{Theorem1} and using (\ref{Lem1ana}), we get
\begin{align}
x_{k-r+1}^{(j)}=x_{k-r+1}+L_{\Gamma_j}A_{k,\Gamma_j}.\label{PThm4c}
\end{align}
Using (\ref{PThm4c}), we derive
\begin{align}
&\!\!\!\!\!\!\!\!x_{k-r+\rho+1}^{(j)}-Ax_{k-r+\rho}^{(j)}-Bu_{k-r+\rho}\nonumber\\
=&x_{k-r+\rho+1}+L_{\Gamma_j}A_{k+\rho,\Gamma_j}-A\big(x_{k-r+\rho}+L_{\Gamma_j}A_{k+\rho-1,\Gamma_j}\big)\nonumber\\
&-Bu_{k-r+\rho}\nonumber\\
%\end{align}
%\begin{align}
=&L_{\Gamma_j}A_{k+\rho,\Gamma_j}-AL_{\Gamma_j}A_{k+\rho-1,\Gamma_j}\label{PThm4d}
\end{align}
where the last equality follows from (\ref{sxk}). From (\ref{PThm4d}), it follows that
\begin{align}
&x_{k-r+\rho+1}^{(j)}-Ax_{k-r+\rho}^{(j)}-Bu_{k-r+\rho}=0 \ is \ equivalent \ to
\nonumber\\
& L_{\Gamma_j}A_{k+\rho,\Gamma_j}-AL_{\Gamma_j}A_{k+\rho-1,\Gamma_j}=0.\label{PThm4d1}
\end{align}
Putting (\ref{PThm4d1}) and Lemma \ref{LSPEa1} together, we prove the lemma.
\par
We now provide a proof of Theorem \ref{Theorem4}. \\
\textit{Sufficiency:}
From (\ref{PThm4c}), it follows that
\begin{align}
x_{k-r+1}^{(v)}\neq x_{k-r+1} \ is \ equivalent \ to
\ L_{\Gamma_v}A_{k,\Gamma_v}\neq0.\label{PThm4e}
\end{align}
Using (\ref{PThm4e}) and noticing $\check{X}_\varsigma(k,r)=\big\{x_{k-r+1}^{(j)}|j\in D_\varsigma\big\}$, we obtain
\begin{align}
&\left.
\begin{array}{c}
v\in D_{\varsigma}\!\!\!\!\!\!\!\!\!\!\!\!\!\!\!\!\!\!\!\!\!\!\!\!\!\!\!\!\!\!\!\!\!\!\!\!\!\!\!\!\!\!\!\!\!\!\!\!\!\!\!\!\!\!\!\!\!\!\!\!\!\!\!\!\!\!\!\!\!\!\!\!\!\!\!\!\!\!\!\!\!\!\!\!\!\!\!\!\!\!\!\!\!\!\!\!\!\\
L_{\Gamma_v}A_{k+\rho,\Gamma_v}-AL_{\Gamma_v}A_{k+\rho-1,\Gamma_v}=0\!\!\!\\
L_{\Gamma_v}A_{k,\Gamma_v}\neq0\!\!\!\!\!\!\!\!\!\!\!\!\!\!\!\!\!\!\!\!\!\!\!\!\!\!\!\!\!\!\!\!\!\!\!\!\!\!\!\!\!\!\!\!\!\!\!\!\!\!\!\!\!\!\!\!\!\!\!\!\!\!\!\!\!\!\!\!\!\!\!\!\!\!\!
\end{array}\right\}\Longrightarrow\left\{
\begin{array}{c}
\!\!\!\!\!\!\!\!\!\!\!\!\!\!\!\!\!\!\!\!v\in D_{\varsigma}\\
\!\!\!L_{\Gamma_v}A_{k,\Gamma_v}\neq0\!\!\!
\end{array}\right\}\nonumber\\
&\ \ \ \ \Longrightarrow\left\{
\begin{array}{c}
\!\!\!x_{k-r+1}^{(v)}\in\check{X}_\varsigma(k,r)\vspace{1pt}\\
\!\!\!x_{k-r+1}^{(v)}\neq x_{k-r+1}
\end{array}\right.\!\!\!.
\nonumber
\end{align}
This completes the proof of the sufficiency.
\\
\textit{Necessity:} Using Lemma \ref{LSPEa2} and noticing $\check{X}_\varsigma(k,r)=\big\{x_{k-r+1}^{(j)}|j\in D_\varsigma\big\}$, we get
the following statement
\begin{align}
& For \ \forall x_{k-r+1}^{(j)}\in\check{X}_\varsigma(k,r),\ it\ holds\ that\  L_{\Gamma_j}A_{k+\rho,\Gamma_j}\nonumber\\
&-AL_{\Gamma_v}A_{k+\rho-1,\Gamma_j}=0.
\label{PThm4g}
\end{align}
If there exists at least a $x_{k-r+1}^{(v)}$ such that $x_{k-r+1}^{(v)}\in \check{X}_\varsigma(k,r)$ and
$x_{k-r+1}^{(v)}\neq x_{k-r+1}$, using (\ref{PThm4e}) and (\ref{PThm4g}), we get
\begin{align}
&\left.
\begin{array}{c}
x_{k-r+1}^{(v)}\in\check{X}_\varsigma(k,r)\vspace{1pt}\!\!\!\\
x_{k-r+1}^{(v)}\neq x_{k-r+1}\!\!\!
\end{array}\right\}\nonumber\\
&\ \ \ \ \ \ \ \ \ \ \Longrightarrow\left\{
\begin{array}{c}
\!\!\!\!\!\!\!\!\!\!\!\!\!\!\!\!\!\!\!\!\!\!\!\!\!\!\!\!\!\!\!\!\!\!\!\!\!\!\!\!\!\!\!\!\!\!\!\!\!\!\!\!\!\!\!\!\!\!\!\!\!\!\!\!\!\!\!\!\!\!\!\!\!\!\!\!\!\!\!\!\!\!\!\!\!\!\!\!\!\!\!\!\!\!\!v\in D_{\varsigma}\\
\!\!\! L_{\Gamma_v}A_{k+\rho,\Gamma_v}-AL_{\Gamma_v}A_{k+\rho-1,\Gamma_v}=0\\
\!\!\!\!\!\!\!\!\!\!\!\!\!\!\!\!\!\!\!\!\!\!\!\!\!\!\!\!\!\!\!\!\!\!\!\!\!\!\!\!\!\!\!\!\!\!\!\!\!\!\!\!\!\!\!\!\!\!\!\!\!\!\!\!\!\!\!\!\!\!\!\!\!\!\!\!L_{\Gamma_v}A_{k,\Gamma_v}\neq0
\end{array}\right.\!\!\!\!\!.
\nonumber
\end{align}
This completes the proof of necessity.

\end{document}